\documentclass[reprint,aps,superscriptaddress,prx,twocolumn]{revtex4-1}
\usepackage{graphicx}% Include figure files
\usepackage{dcolumn}% Align table columns on decimal point
\usepackage{bm}% bold math
\usepackage{amsmath}% http://ctan.org/pkg/amsmath
\usepackage{amssymb}%花体字母加粗
\usepackage{mathrsfs}%花体字母
\usepackage{amsmath,amssymb,amsthm}
\usepackage{braket}
\usepackage[utf8]{inputenc}
\usepackage{tikz}
\usetikzlibrary{matrix,shapes,arrows,positioning,chains}
\usetikzlibrary{calc}
\usepackage{etoolbox}
\usepackage[hyperindex,breaklinks]{hyperref}% add hypertext capabilities

\hypersetup{colorlinks=true, citecolor=blue, urlcolor=blue, linkcolor=blue}
\newtheorem{thm}{Theorem}
\newtheorem{definition}{Definition}
\DeclareMathOperator{\Tr}{Tr}
\begin{document}

% Use the 'preprinting' class option to override journal defaults
%\preprint{}

\title{Quantum observation scheme universally identifying causalities from correlations}

% address or URL should go in the {}'s for \email and \homepage.

\author{Chenguang Zhang}
\affiliation{College of Intelligence and Computing(School of Computer Science and Technology), Tianjin University, No.135, Ya Guan Road, Tianjin, 300350, China.}
\affiliation{School of Science, Hainan University, No. 58, Renmin Avenue, Haikou, 570228, China.}

\author{Yuexian Hou}
\email[Corresponding author: ]{yxhou@tju.edu.cn}
%\thanks{}
\affiliation{College of Intelligence and Computing(School of Computer Science and Technology), Tianjin University, No.135, Ya Guan Road, Tianjin, 300350, China.}

\author{Dawei Song}
%\thanks{}
\affiliation{Faculty of Science, Technology, Engineering and Mathematics College of Intelligence and Computing, The Open University, Walton Hall, Milton Keynes, MK7 6AA, United Kingdom.}

%Correspondence and requests for materials should be addressed to Y.H.

\date{\today}

\begin{abstract}
It has long been recognized as a difficult problem to determine whether the observed statistical correlations between two classical variables arise from causality or from common causes.
Recent research has shown that in a quantum theoretical framework, the mechanisms of entanglement and quantum coherence provide advantages in tackling this problem.
In some particular cases, quantum common causes and quantum causality can be effectively distinguished by using observations alone.
However, these solutions do not apply to all cases. There still exist a large class of cases in which quantum common causes and quantum causality cannot be distinguished.
In this paper, along the line of considering unitary transformation as causality in the quantum world, we formally show that quantum common causes and quantum causality are universally separable. Based on the analysis, we further provide a general method to discriminate the two.
\end{abstract}

% insert suggested keywords - ASP authors don't need to do this
%\keywords{}

\maketitle

\section{INTRODUCTION}
Common causes and causality are two building blocks in the Reichenbach's principle of causal explanation \cite{renoirte1956hans}. This principle asserts that if two observed variables are found to be statistically correlated, it is possible that the early variable directly causes the later one, i.e., the causality case, or that the two share a common cause, i.e., a correlation between them. In this paper, we focus on identifying the causality from the correlations in the quantum world using only experimental observations.

Despite the central role of causal explanations in science, discriminating causality from correlations is still a nontrivial issue. In classic cases, it is only recently that a rigorous framework for causal inference has been developed \cite{pearl2009causality}. Its core ingredient is the possibility of external interventions on the early variable. For example, in a drug trial, randomizing the assignment of drug or placebo as the intervention is the key step to detect the potential causality between the treatment and recovery.  

In quantum cases, the Bell theorem rules out the classical common cause explanation of the causal models that obey the Bell inequality \cite{wood2015lesson}. Considerable efforts have recently been devoted to make causal models compatible with the quantum mechanism, including applying the classical causal model by introducing hidden and fine tuned mechanisms \cite{evans2012new} or alternatively transferring classical causal modeling tools to the quantum domain \cite{tucci1995quantum, laskey2007quantum, pienaar2015graph, fritz2016beyond, leifer2006quantum, henson2014theory}, thus leading to a reformulation of quantum causal models \cite{allen2017quantum,giarmatzi2018quantum, costa2016quantum}. 
Causal structures are usually represented as directed acyclic graphs in these methods and the established quantum version of Reichenbach's principle allows one to perform Bayesian inference to analyze the causal structures.
%Since quantum mechanism can be seen as a theory about interventions, these methods in this sense can be regarded as natural extensions of classical %interventionist schemes in the quantum domain \cite{costa2016quantum} and the established quantum version of Reichenbach's principle allows one to %perform Bayesian inference to explore the causal structures. % \cite{allen2017quantum,giarmatzi2018quantum}.

%%In contrast with these methods, As another aspect of causal inference, our work focuses on quantum observational scheme,
In contrast with these methods, our work is a quantum observational scheme, where in analogy to the classical observational scheme, only observations, namely, the (local) projective measurements in arbitrary orthogonal bases, are allowed; however, general interventions, such as the unitary transformation on the quantum state and the state preparation, are forbidden. Note that the quantum observational mechanism is of the relevance since, in the contexts of quantum causality discovery, the intervention effect implemented by quantum measurement is considerably limited and should be differentiated from a real quantum intervention mechanism in analogy with the classical intervention mechanism. Also note that, unlike the case of the interventionist scheme, justifying an operational solution for quantum observational schemes has no strict classical analogy since, without additional assumptions, its classical version is actually impossible. Exploring the operational quantum observational schemes and clarifying their potential quantum advantages would help to advance a better conceptual and technological understanding of quantum causality discovery.

Research on the quantum observational scheme can date back to the work of Fitzsimons et al., where an irregular pseudo-density operator was defined as a witness of causality \cite{fitzsimons2013quantum}. Furthermore, Ried et al. developed this work and justified that a passive observational scheme is sufficient for distinguishing the common causes (the quantum states) from the direct causes (the quantum channels) in some extreme cases \cite{ried2015quantum}. However, to obtain the same state as before the measurement, their passive observational scheme was defined to require the common causes only as the locally maximally mixed states, which limits the application scope of the scheme and would actually be inappropriate since whether a procedure belongs to the observational scheme or not should be intrinsically defined rather than depend on the input objects.

Our work generalizes the passive observational scheme, where the common causes are allowed to be any density operator and the projective measurements are no longer restricted in a fixed basis but could be done in arbitrary orthogonal bases. This scheme equals the active quantum observation scheme in Ref. \cite{kubler2018two}.
Additionally, in the present work we focus on this scheme, but only consider the unitary channels as the direct causes and exclude other completely positive trace preserving (CPTP) channels for simplicity and conceptual clarity since the mixed mechanisms in the CPTP channels may make different CPTP maps indistinguishable by observation alone \cite{chiribella2010probabilistic}.
%are no longer restricted in a fixed basis but could be in arbitrary orthogonal bases
%observations are no longer restricted as the projective measurements in a fixed basis but could be the projective measurements in arbitrary orthogonal bases

Our work then aims to put forward a unified and practical method to completely distinguish the direct cause from the common cause in the general cases. Ried et al. showed that if the common causes are considered as the maximally entangled states, a complete solution of causal inference by observational scheme is possible \cite{ried2015quantum}. However, when the general scheme is considered, the existing methods \cite{hu2018discrimination, kubler2018two} may fail to distinguish the common cause from the direct cause because their observation results, e.g., the vector-valued statistics $\bold{P}$ in Ref. \cite{hu2018discrimination} or the signed singular values (SSV) in Ref. \cite{kubler2018two} of common cause and direct cause, could be the same (see Fig. \ref{fig1}). Additionally, although Ried et al. and Kübler et al. argued that the causal inference could benefit from the signaling when the general common causes are considered \cite{ried2015quantum, kubler2018two}, the causal inference problem actually cannot be solved by signaling because some input states disable the signaling, and we may have no prior knowledge about the current input states to determine whether the signaling is usable or not. Solving these problems would soundly indicate the quantum advantage in a general scope and would also extend the potential application scope of the observational scheme, which may be important for the applications that allow the input to be any density operator, e.g., the non-Markovianity testing \cite{rivas2014quantum, laine2010measure} in which the environment acts as a common cause of the system and the quantum gate discrimination \cite{chiribella2013identification,chiribella2012perfect}.%chiribella2012perfect}.

In the present work we follow the same setup as in Ref. \cite{hu2018discrimination} and consider the quantum system with two temporally ordered qubits. We first analyze the possible quantum common causes and quantum direct causes in terms of a vector-valued statistics $\bold{P}$ and discuss how the $\bold{P}$ changes when the unitary operators are applied to the observables from which $\bold{P}$ is derived. Second, aiming at the overlapping area in which the quantum common causes and quantum direct causes have the same $\bold{P}$ values, thereby being indistinguishable (see Fig. \ref{fig1}), we show how to design appropriate unitary operators to bring $\bold{P}$ values of possible direct causes out of the overlapping area, based on which we then prove that the quantum common causes and the quantum direct causes could be distinguished. Particularly, we find that for some cases with $\bold{P}=(0,0,1)'$, the unitary operators applied to the observables of one side of the system are necessarily distinct from those in another side to promise an operational solution (the symbol ``$ ' $ '' represents the conjugate transpose throughout the paper). Finally, a general identification method is given. Simulation experiments verify our theoretical results.
.
\section{Conduct Unitary Operations with Possible Quantum Common Causes and Quantum Direct Causes}
\subsection{Possible quantum common causes and quantum direct causes}
\begin{figure}
\centering
\includegraphics[width=3in,height=2in]{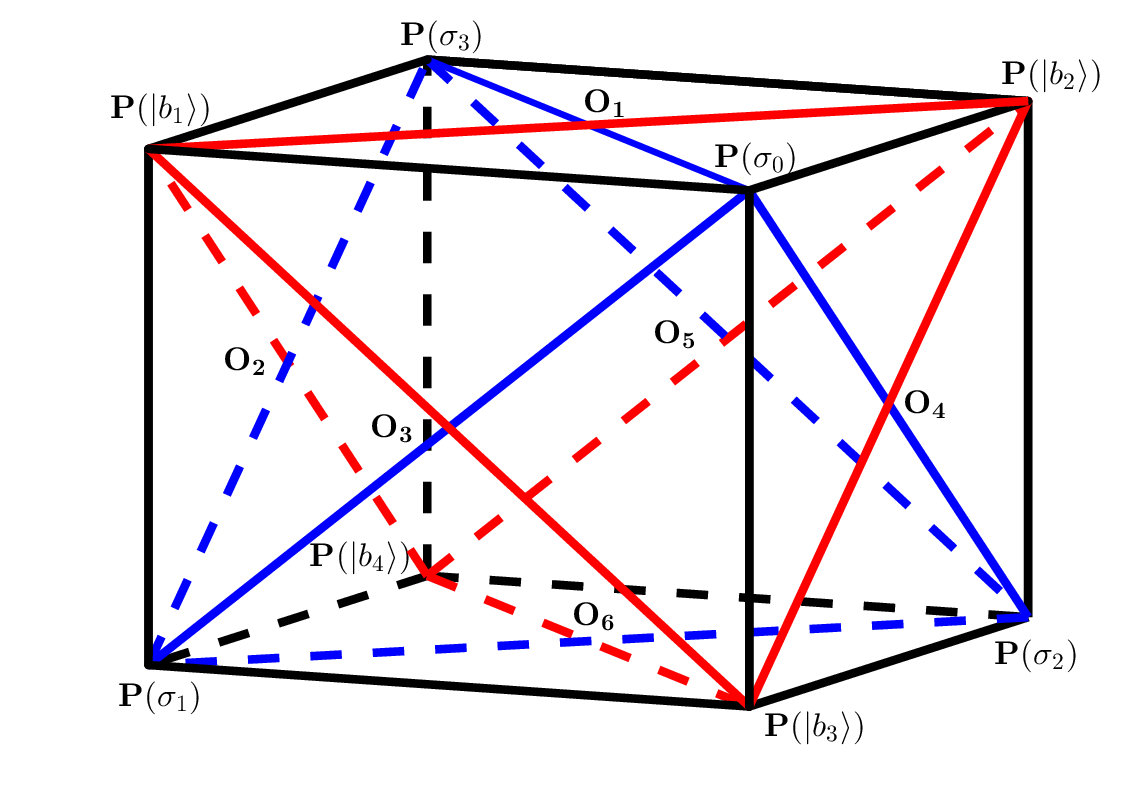}
\caption{ The geometric interpretation of the statistic $\bold{P}$. If the correlation of two qubits is arose from quantum common causes, $\bold{P}$ should lie in the red regular tetrahedron(denoted by TCC) with vertices $\bold{P}(\ket{b_1})=(1,-1,1)'$, $\bold{P}(\ket{b_2})=(-1,1,1)'$, $\bold{P}(\ket{b_3}=(1,1,-1)'$ and $\bold{P}(\ket{b_4})=(-1,-1,-1)'$. The four vertices form a basis for $\bold{P}$ in TCC. If the correlation is arose from a direct cause between the two qubits, $\bold{P}$ should lie in the blue regular tetrahedron(denoted by TDC) with vertices $\bold{P}(\sigma_0)=(1,1,1){'}$,$\bold{P}(\sigma_1)=(1,-1,-1){'}$, $\bold{P}(\sigma_2)=(-1,1,-1){'}$ and $\bold{P}(\sigma_3)=(-1,-1,1){'}$. The four vertices form a basis for $\bold{P}$ in TDC. Obviously, TCC and TDC have an overlapping area, in which quantum common causes and quantum causality are indistinguishable.}
\label{fig1}
\end{figure}

We review the vector-valued function $\bold{P}$ as well as its related properties in \citep{hu2018discrimination} first (see Fig. \ref{fig1}). Given a two-qubit system represented by a density operator $\rho$, we measure these two qubits with the same one of three Pauli observables $\sigma_i(i=1,2,3)$ respectively and assume the outcomes are $k$ and $m$ respectively. Then define
\begin{equation}\label{cdef}
   C_{ii}(\rho)=p(k=m|ii)-p(k\neq m|ii)
\end{equation}
and
\begin{equation}\label{Pdef}
\bold{P}(\rho)=
  \begin{pmatrix}
      C_{11}(\rho) \\ C_{22}(\rho) \\ C_{33}(\rho)
  \end{pmatrix}.
\end{equation}
When $\rho$ represents an entangled state or a correlated mixture of separable states, it is a common cause. Specially, if $\rho$ is a pure state identified with $\ket{\phi}$ and has a representation in terms of Bell states, i.e., $\ket{\phi} = \sum_{i=1}^{4}w_i\ket{b_i}$, where $\sum_{i=1}^{4}w_i^2=1, w_i \in \mathbb{R}$ and $\ket{b_i}(i=1,2,3,4)$ is one of the four Bell states, then
\begin{equation}\label{cdecomp}
\bold{P}(\ket{\phi})=\sum_{i=1}^{4}w_i^2 \bold{P}(\ket{b_i}).
\end{equation}
Except for the quantum common cause, quantum causality is also a possible explanation of the observed quantum correlation. In this case, there is a unitary transformation $\bold{U}$, i.e., a direct cause, between the measured states of the two qubits(which are actually the same qubit sequentially occurring twice). As in the common cause case, the same measurements are take on the qubit before and after the transformation $\bold{U}$, to get the statistic $\bold{P}$. It was proven in Ref. \citep{hu2018discrimination} that the $\bold{P}$ value in this case does not depend on the state of the early qubit, but on $\bold{U}$. Then $\bold{P}$ can be regarded as a function of $\bold{U}$. And we denote it by $\bold{P}(\bold{U})$. For any given $\bold{U}$, it was showed that there exist $p_j \geq 0$ satisfying $\sum_{j=0}^{3}p_j=1$ such that
%$\bold{P}$ value is obtained by measuring one of the three Pauli observables $\sigma_i(i=1,2,3)$ on the qubit before and after the transformation $U$. Since the $\bold{P}$ value of this system does not depends on the state of the early qubit,
\begin{equation}\label{udecomp}
\bold{P}(\bold{U})=\sum_{j=0}^{3}p_j \bold{P}(\sigma_j),
\end{equation}
where $\sigma_j(j=0,1,2,3)$ is one of the four Pauli matrices(including the identity matrix $\sigma_0$).

The value of $\bold{P}$ can be used to evaluate the existence of quantum causality. However, as stated in the introduction, when the value of $\bold{P}$ is in the overlapping area, more designed measurements are needed. To this end, we first analyze the current measurement result, which is represented as Eq. \eqref{cdecomp} with $w_i(i=1,2,3,4)$ or Eq. \eqref{udecomp} with $p_i(i=0,1,2,3)$, to get the general representation forms of possible quantum common causes and possible quantum direct causes. We show them in Lemma \ref{lemma1} and \ref{lemma2}.
\newtheorem{Lemma}{Lemma}
\begin{Lemma}\label{lemma1}
Given $w_j(j=1,2,3,4)\in \mathbb{R}$ satisfying $\sum_{j=1}^{4}{w_j^2}=1$, if only pure states are considered, there is a unique family of states 
\begin{equation}\label{eq5}
    \ket{\phi}=\sum_{j=1}^{4}w_je^{i\theta_j}\ket{b_j},
\end{equation}
in the parameters $\theta_j$, such that $\bold{P}(\ket{\phi})= \sum_{j=1}^{4}{w_j^2 \bold{P}(\ket{b_j})}$, where $\theta_j$ called the phase of $w_j$ can be any value in $[0, 2\pi)$. The set of all the pure states above with the same $\bold{P}(\ket{\phi})$ value denotes by $\Phi(w_1, w_2, w_3, w_4)$.
\end{Lemma}
The proof is in the Appendix \ref{pl1}.

Obviously, if mixed quantum states as common causes are considered, the mixed quantum states represented as a convex combination of the pure states in Lemma \ref{lemma1} can also meet the requirement of Lemma \ref{lemma1}.
\begin{Lemma}\label{lemma2}
Given $p_j\geq 0 (j=0,1,2,3) $ satisfying $\sum_{j=0}^{3}{p_j}=1$, there are 16
\begin{equation}\label{Uform}
U=e^{i\frac{\theta}{2}}
\begin{pmatrix}
  e^{i \gamma_1 }cos(\varphi_0) & e^{i\gamma_2}sin(\varphi_0)\\
  -e^{-i\gamma_2}sin(\varphi_0) & e^{-i\gamma_1}cos(\varphi_0)
\end{pmatrix},
\end{equation}
up to the global phase $\frac{\theta}{2} \in [0,\pi)$, such that $\bold{P}(U)= \sum_{j=0}^{3}{p_j\bold{P}(\sigma_j)}$, where
$cos(\varphi_0) = \sqrt{c_1}$, $sin(\varphi_0) = \sqrt{d_1}$,
$\gamma_1=(-1)^{n_1}{\arccos(\frac{c_2}{c_1})}{/2}  + k_1\pi$(if $c_1 = 0$, let $\gamma_1=0$),
$\gamma_2=(-1)^{n_2}{\arccos(\frac{d_2}{d_1})}{/2}  + k_2\pi$(if $d_1 = 0$, let $\gamma_2=0$),
$c_1 = p_0 + p_3$, $c_2 = p_0 - p_3$, $d_1 = p_1 + p_2$, $d_2 = p_2 - p_1$, 
and $n_1, n_2, k_1, k_2 \in \{0, 1\}$. The set of all the above unitary matrices with the same $\bold{P}(U)$ value denotes by $\mathcal{U}(p_0, p_1, p_2, p_3)$.
\end{Lemma}
%U = [sqrt(c1)*exp(1i *(gamm1 + theta)/2),  sqrt(d1) * exp(1i *(gamm2 + theta)/2);
%                         -sqrt(d1)*exp(1i *(-gamm2 + theta)/2), sqrt(c1)*exp(1i *(-gamm1 + theta)/2)];
%
The proof is in Appendix \ref{pl2}.
\subsection{The changes of $\bold{P}$ when the observables are transformed}
Based on the analysis results, by applying unitary operators $V$ on the observables of the two qubits respectively, we are interested in whether there are differences between the changes of the value of $\bold{P}$ of the common cause case and that of the causality case, where ${V} \in \bold{U}(2)$ (whose global phase is omitted) is expressed as
\begin{equation}\label{V}
 V=
 \begin{pmatrix}
  e^{i \psi}cos(\varphi) & e^{i\chi}sin(\varphi)\\
  -e^{-i\chi}sin(\varphi) & e^{-i\psi}cos(\varphi)
\end{pmatrix}.
\end{equation}
These differences may diverge quantum common causes and quantum causality, which is the starting point of our subsequent analysis. We introduce the following definition firstly:
\begin{definition}%(or two qubits with a unitary transformation $U \in U(2)$ between them)
Given two qubits represented by a density operator $\rho$ and a unitary operator $V$, measuring the observables $V\sigma_iV'(i=1,2,3)$ on the two qubits respectively gives new values of $\bold{P}(\rho)$, $C_{ii}(\rho)$ and the probabilities $p(k=m|ii)$ as well as $p(k\neq m|ii)$. Denote them by $\bold{P}_V(\rho)$, $C_{ii\_V}(\rho)$ and $p_{_V}(k=m|ii)$ as well as $p_{_V}(k\neq m|ii)$ respectively.
\end{definition}
%It have been shown that $\bold{P}'(U)=\bold{P}'(V'UV)$.
We first discuss the common cause case. Over the set of possible quantum common causes, the general calculation formula of $\bold{P}_V$ for any unitary operator $V$ is shown in the following Lemma \ref{lemma3}.  %of Lemma \ref{lemma3} and its
\begin{Lemma}\label{lemma3}
Quantum common causes scenario. Given two qubits in the quantum state $\rho$, for any unitary operator $V \in \bold{U}(2)$ as stated in Eq. \eqref{V},
\begin{equation}
  p_{_V}(k=m|ii)=\Tr\left(\left(V \otimes V \right) \left(\xi_{i}^{dd} + \xi_{i}^{uu}\right) \left(V \otimes V \right){'}\rho\right).
\end{equation}
In particular, if $\rho=\ket{\phi}\bra{\phi}$ and $\ket{\phi}$ is a quantum pure state with $\ket{\phi}=\sum_{j=1}^{4}w_je^{i\theta_j}\ket{b_j}$ as stated in Eq. \eqref{eq5}, thus
\begin{equation}
\begin{aligned}
  &p_{_V}(k=m|ii)\\
  &= \Tr\left(B{'}\left(V \otimes V \right) \left(\xi_{i}^{dd} + \xi_{i}^{uu}\right) \left(V \otimes V \right){'} Bww{'}\right),
\end{aligned}
\end{equation}
where $\xi_{i}^{uu}$ and $\xi_{i}^{dd}$ are spectral measures associated with the observable $\sigma_i\otimes\sigma_i(i=1,2,3)$, reflecting whether both qubits are pointing the same direction, $B=\left (\ket{b_1}, \ket{b_2}, \ket{b_3}, \ket{b_4}\right )$ and
$w = \begin{pmatrix}
  w_1e^{-i\theta_1}, &  w_2e^{-i\theta_2}, & w_3e^{-i\theta_3}, & w_4e^{-i\theta_4}
  \end{pmatrix}{'}$.
\end{Lemma}
 %and the set of possible quantum causalities with an arbitrary unitary operator $V$, and Lemma \ref{lemma4} respectively. In addition, some corollaries are given. They are useful for the design of the appropriate unitary
%operators. We discuss the design of the unitary operator in the next section.
\newtheorem{corollary}{Corollary}[Lemma]
%$.Because the computation of $C_{33\_V}$ is relatively simple,$C_{33\_V}$ plays an important role in our work.
The proof is in the Appendix \ref{pl3}.

\begin{figure}
\centering
\includegraphics[width=3.0in,height=2in]{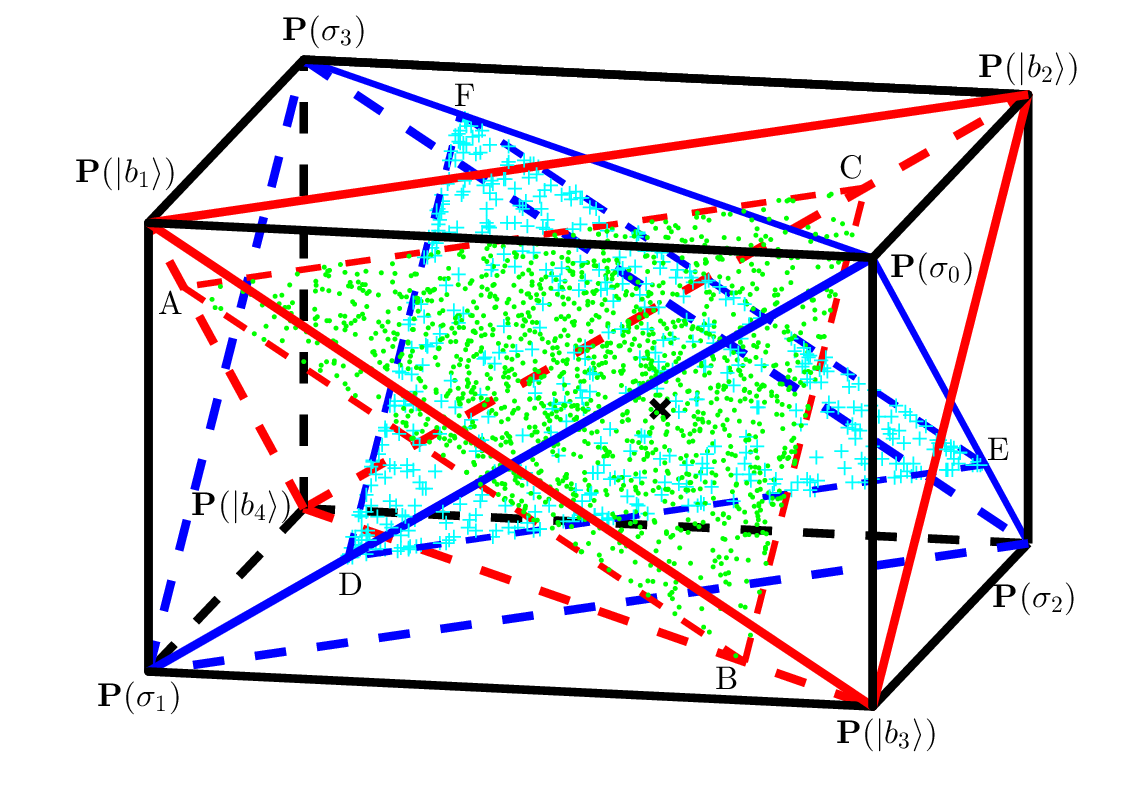}
\caption{The geometric interpretation of the range of the value of $\bold{P}_V$. Given an initial value of $\bold{P}$(represented by the black symbol ``$\times$") in the plane $l(b)$, analyze the current value of $\bold{P}$ to get a possible common cause $\rho$ and a possible direct cause $U$. Then with some arbitrary unitary operations $V$, the values of $\bold{P}_V(\rho)$(represented by the symbol ``.") lie in the intersection area of TCC and the plane $l(b)$, i.e., the triangle with vertices A,B and C; the values of $\bold{P}_V(U)$(represented by the symbol ``+") lie in the intersection area of TDC and the plane $l(b)$, i.e., the triangle with vertices D, E and F.}
\label{figf2}
\end{figure}

As a special case of Lemma \ref{lemma3}, we give the following corollary for the computation of $C_{33\_V}$ on particular pure states with $w_4=0$ since our many following works are in large part associated with the analysis of the changes of the value of $C_{33\_V}$.
\begin{corollary}\label{c31}
Specially, given $\ket{\phi}=\sum_{j=1}^{4}w_je^{i\theta_j}\ket{b_j}$ with $w_4=0$, after having applied the unitary operation $V$ on the observables of the two qubits respectively, We have
\begin{equation}\label{eqc31}
\begin{aligned}
&C_{33\_V}(\ket{\phi})=C_{33}(\left(V \otimes V \right)'\ket{\phi})=2p_{_V}(k=m|33) - 1\\
& = 2\Tr\left({\begin{pmatrix}
      1 - \tau_4     & -\tau_1 & -\tau_2 & 0 \\
      \tau_1                                   & 1- \tau_5 & \tau_3 & 0 \\
      \tau_2 & \tau_3 & \frac{1}{2}\left(1 - \cos(4\varphi)\right) & 0 \\
      0 & 0 & 0 & 0
    \end{pmatrix}}ww'\right) - 1,
\end{aligned}
\end{equation}
where $\tau_1 = \frac{i}{2}\sin^2(2\varphi)\sin(2\chi + 2\psi)$, $\tau_2 = \frac{i}{2}\sin(4\varphi)\sin(\chi + \psi)$,
$\tau_3 = -\frac{1}{2}\sin(4\varphi)\cos(\chi + \psi)$, $\tau_4 = \sin^2(2\varphi)\sin^2(\chi + \psi)$, $\tau_5 =  \sin^2(2\varphi)\cos^2(\chi + \psi)$. Obviously, the value of $C_{33}(\left(V \otimes V \right)'\ket{\phi})$ does not depend on the respective value of $\chi$ or $\psi$ but on the sum of them.
\end{corollary}
%We find that though different quantum common cause behind the initial $\bold{P}$ or different unitary evaluation matrix applied on the observables may lead to the different value of $\bold{P}_V$,
According to the above results, with different possible quantum common causes behind the same $\bold{P}$ or different $V$, the values of $\bold{P}_V$ are usually different. But as shown in the following Corollary \ref{c32},
we find that these values have some degree of consistency and are always on the same plane, which is determined by the initial value of $\bold{P}$ (see Fig. \ref{figf2}).
\begin{corollary}\label{c32}
Given $w_j \in \mathbb{R}(j=1,2,3,4)$ as stated in Lemma \ref{lemma1}, $\forall\ket{\phi}\in\Phi(w_1,w_2,w_3, w_4)$ and $\forall V \in \bold{U}(2)$ as stated in Eq. \eqref{V}, $\bold{P}_V(\ket{\phi})$ takes values in a fixed plane. The plane is determined only by $\{w_j|j=1,2,3,4\}$ and is independent of the choice of $V$ and the phase of $w_j(j=1,2,3,4)$.
\end{corollary}
The proof is in the Appendix \ref{pc32}.

In the quantum causality scenario, similar analyses are done. We first get the general calculation formula of $\bold{P}_V$ over the set of possible quantum direct causes for any unitary matrix $V$. We show it in Lemma \ref{lemma4}. Also we prove in Corollary \ref{c41} that the set of possible quantum direct causes can be divided into four subsets in terms of the values of $\bold{P}_V$ over them. This corollary is useful since it reduces the number of unitary operators that we need to deal with.
\begin{Lemma}\label{lemma4}
Quantum causality scenario. Given $U \in \mathcal{U}(p_0, p_1, p_2, p_3)$ as stated in Eq. \eqref{Uform} and $V \in \bold{U}(2)$ as stated in Eq. \eqref{V}, applying V in temporal order on the observables of the two qubits respectively, then $\bold{P}_V(U)=\bold{P}(V'UV)$. Furthermore, $C_{11\_V}(U)=2(a_1^2 + b_2^2)-1$, $C_{22\_V}(U)=2(a_1^2 + b_1^2)-1$ and $C_{33\_V}(U)=2(a_1^2 + a_2^2)-1$, wherein
%\small{
%\begin{equation}
%\begin{aligned}
%% \nonumber % Remove numbering (before each equation)
% &a_1=\cos(\varphi_0)\cos(\gamma_1) \\
% &a_2=\cos(\varphi_0)\cos(2\varphi)\sin(\gamma_1)-\sin(\varphi_0)\sin(2\varphi)\sin(\gamma_2-\psi-\chi)\\
% &b_1=\sin(\varphi_0)\cos^2(\varphi)\cos(\gamma_2-2\psi)+\\ &\sin(\varphi_0)\sin^2(\varphi)\cos(\gamma_2-2\chi) - \cos(\varphi_0)\sin(2\varphi)\sin(\chi-\psi)\sin(\gamma_1)\\
% &b_2=-\sin(\varphi_0)\sin^2(\varphi)\sin(\gamma_2-2\chi)+ \sin(\varphi_0)\cos^2(\varphi)\sin(\gamma_2-2\psi)\\
% &+\cos(\varphi_0)\sin(2\varphi)\cos(\chi-\psi)\sin(\gamma_1).
%\end{aligned}
%\end{equation}
%}
$a_1=\cos(\varphi_0)\cos(\gamma_1)$, $a_2=\cos(\varphi_0)\cos(2\varphi)\sin(\gamma_1)
-\sin(\varphi_0)\sin(2\varphi)\sin(\gamma_2-\psi-\chi)$, $b_1=\sin(\varphi_0)\cos^2(\varphi)\cos(\gamma_2-2\psi)+ \sin(\varphi_0)\sin^2(\varphi)\cos(\gamma_2-2\chi)- \cos(\varphi_0)\sin(2\varphi)\sin(\chi-\psi)\sin(\gamma_1)$, $b_2=-\sin(\varphi_0)\sin^2(\varphi)\sin(\gamma_2-2\chi)+ \sin(\varphi_0)\cos^2(\varphi)\sin(\gamma_2-2\psi)
 +\cos(\varphi_0)\sin(2\varphi)\cos(\chi-\psi)\sin(\gamma_1)$.
\end{Lemma}
The proof is in the Appendix \ref{pl4}.
\begin{corollary}\label{c41}
Given $p_j\geq 0 (j=0,1,2,3)$ as stated in Lemma \ref{lemma2} and $V \in \bold{U}(2)$, the image set of $\bold{P}_V(U)$ $($$U \in \mathcal{U}(p_0, p_1, p_2, p_3)$$)$ contains at most four different elements.
\end{corollary}
The proof is in Appendix \ref{pc41}.

Next, just like the quantum common cause case, we find for any possible direct cause behind the initial value of $\bold{P}$ and for any unitary operator $V$, $\bold{P}_V$ also lie in a fixed plane(see Fig. \ref{figf2}).
\begin{corollary}\label{c42}
Given $p_j\geq 0 (j=0,1,2,3)$ as stated in Lemma \ref{lemma2}, $\forall U\in \mathcal{U}(p_0, p_1, p_2, p_3)$ and $\forall V\in \bold{U}(2)$ as stated in Eq. \eqref{V}, $\bold{P}_V(U)$ takes values in a fixed plane. The plane is determined only by $\{p_j|j=0,1,2,3\}$ and is independent of the choice of $V$.
\end{corollary}
The proof is in the Appendix \ref{pc42}.

Finally, we show that the plane mentioned in Corollary \ref{c32} is identical to the plane stated in Corollary \ref{c42} when the value of $\bold{P}$ discussed in the two corollaries are the same. This property motivates us to discuss the discrimination problem plane by plane(see the next section). We summarize the corresponding results as Lemma \ref{lemma5} and Lemma \ref{lemma6}, wherein the quantum-common-cause part of Lemma \ref{lemma5} only discusses the pure quantum states and Lemma \ref{lemma6} discusses the general case, i.e., general quantum common causes including mixed quantum states.
\begin{Lemma}\label{lemma5}
Analyzing the initial given value of $\bold{P}$ to get $\Phi(w_1,w_2,w_3, w_4)$ and $\mathcal{U}(p_0, p_1, p_2, p_3)$, then $\forall V\in \bold{U}(2)$, $\bold{P}_V(\Phi)$ and $\bold{P}_V(\mathcal{U})$ are always lying in a same plane, where $\bold{P}(\Phi)$ and $\bold{P}(\mathcal{U})$ are respectively the image sets of $\bold{P}$ over the set $\Phi$ and $\mathcal{U}$. The plane's normal vector is $(1,1,1)'$ and its constant term ranges from -1 to 1.
\end{Lemma}
The proof is in Appendix \ref{pl5}.
\begin{definition}
We denote the constant term $4p_0 - 1$ or $1 - 4w_4^2$ by $b$. And since these planes differ from each other only by the constant term, we use $l(b)$ to represent the plane with the constant term $b$.
\end{definition}
\begin{Lemma}\label{lemma6} %after having applied an arbitrary unitary operation on the observables of the two qubits respectively in temporal order
Given an initial measurement result of $\bold{P}$ of two qubits in the plane $l(b)$, with any unitary operator $V$, for any possible common cause $\rho$ and for any possible direct cause $U$, $\bold{P}_V(\rho)$, and $\bold{P}_V(U)$ are still in the plane $l(b)$(see Fig. \ref{figf2}).
\end{Lemma}
The proof is in the Appendix \ref{pl6}.
\section{Design of unitary operators}
In this section, we show how to design unitary operators to get appropriate $\bold{P}_V$ functions for the discrimination task. It can be seen from Lemmas \ref{lemma5} and \ref{lemma6} that no matter what unitary matrix is chosen, $\bold{P}_V$ is always in the $l(b)$ plane that $\bold{P}$ is initially in. This prompts us to take the area that is in the plane and in which the respective value of $\bold{P}_V$ of quantum common causes and quantum direct causes do not overlap as the target of $\bold{P}_V$.

Compared with the difficulty of handling the infinite cases of possible quantum common causes, it is relatively easy to deal with the possible 16 cases of quantum causality(see Theorem \ref{th1} below). Furthermore, by Lemma \ref{lemma4}, $C_{33\_V}$ is formally simpler than $C_{11\_V}$ and $C_{22\_V}$. And we notice that given $\bold{P}$ in the plane $l(b)(b\neq 1)$, among the points that belong to the image set of $\bold{P}_V(U)(U\in \mathcal{U})$, $\bold{P}_V$ with $C_{33\_V}$ being 1 is one of the possible points that are farthest from the image set of $\bold{P}_V(\ket{\phi})(\ket{\phi}\in \Phi)$(see Fig. \ref{figf2}). Based on the above considerations, the design of unitary operators aims to transfer the third entry of $\bold{P}_V$, i.e., $C_{33\_V}$ to 1 when there is a causality between the two qubits. To implement this idea, two questions need to be answered. The first question is that, given an initial value of $\bold{P}$, whether there are appropriate operators $V$ such that for all possible cases of quantum causality, $C_{33\_V}$ are equal to 1? The second question is whether we can conclude there exists a quantum causality when $C_{33\_V}$ is equal to 1?

\begin{figure}
\centering
\includegraphics[width=3in,height=2in]{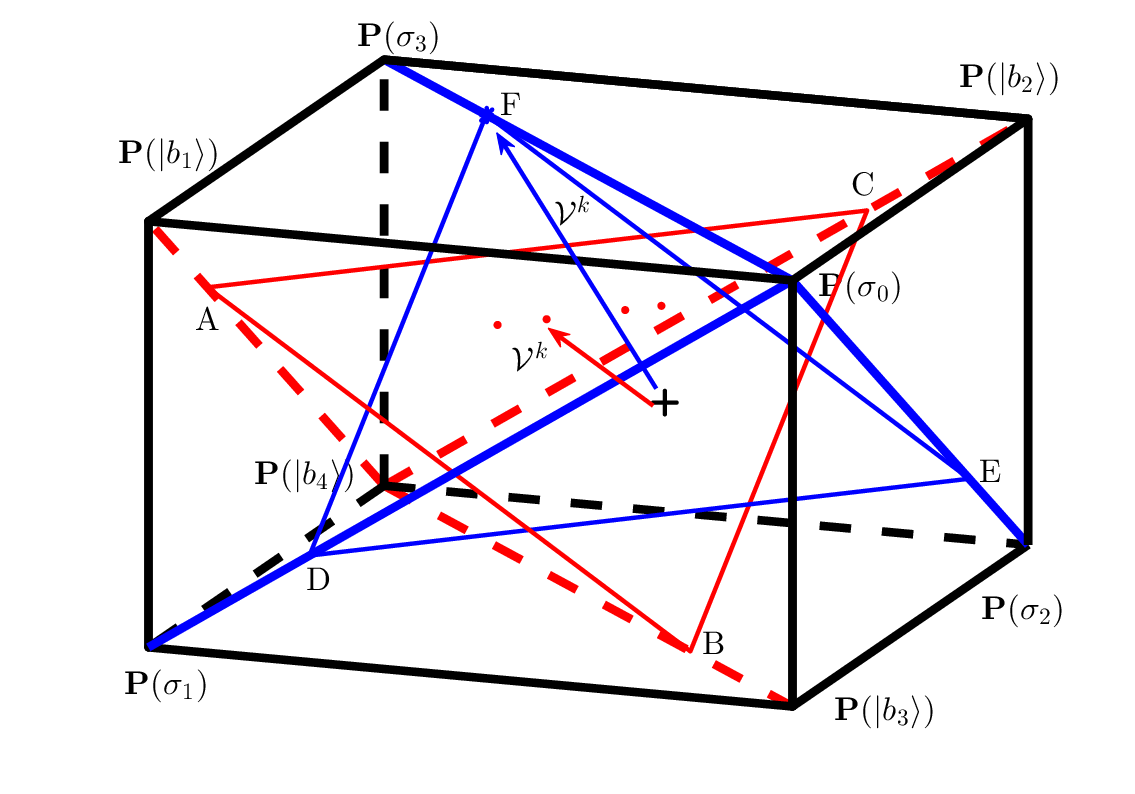}
\caption{The cases where $\bold{P}$ is initially in $l(b)(b \neq 1)$. Given the statistic $\bold{P}$(represented by the black symbol ``+'') of two qubits in the plane $l(b)(b\neq 1)$, if there is a direct cause $U\in \mathcal{U}_k(k=1,2,3,4)$ between the two qubits, then $\forall V \in \mathcal{V}^k$, $\bold{P}_V$ is transferred to the point F with third entry being 1. However, if the two qubits have a common cause acting on them, then $\forall k(k=1,2,3,4)$ and $\forall V \in \mathcal{V}^k$, $\bold{P}_V$(represented by the red symbol ``.'') is not equal to the point F.}
\label{fig3}
\end{figure}

As presented in Corollary $\ref{c41}$, given $p_j \geq  0(j=0,1,2,3)$, $\mathcal{U}(p_0, p_1, p_2, p_3)$ can be divided into four subsets according to the values of $\bold{P}_V$ on them. The four subsets denotes by $\mathcal{U}_k(p_0, p_1, p_2, p_3)(k=1,2,3,4)$. For the first question, we first prove that with carefully designed unitary operators acting on the observables, the third entry of $\bold{P}_V(\mathcal{U}_k)(k=1,2,3,4)$ can be equal to 1(see Fig. \ref{fig3}). For the second question, we prove for any possible quantum common cause $\rho$, with any unitary operator $V$, the entries of $\bold{P}_V(\rho)$ can not be equal to 1, unless $\bold{P}(\rho)$ is initially in the plane $l(1)$(see Fig. \ref{fig3}). The results are shown in Theorem \ref{th1} and Theorem \ref{th2}.
\begin{thm}\label{th1}
Given $p_j(j=0,1,2,3)$ and $k\in\{1,2,3,4\}$, $\forall U \in \mathcal{U}_k(p_0, p_1, p_2, p_3)$, there exist unitary operators $V$ as stated in Eq. \eqref{V} with
\begin{equation}\label{th11}
  \psi+\chi = \gamma_2-\frac{k_1\pi}{2}, \varphi = \frac{k_2\pi - \omega}{2},
\end{equation}
such that $\bold{P}_V(U)=(2p_0-1,2p_0-1,1){'}$, where $\sin(\omega)= \frac{\sin(\varphi_0)\sin(\gamma_2-\psi-\chi)}{r}$, $\cos(\omega) = \frac{\cos(\varphi_0)\sin(\gamma_1)}{r}$ and
$r = [\cos^2(\varphi_0)\sin^2(\gamma_1) + \sin^2(\varphi_0)\sin^2(\gamma_2-\psi-\chi)]^{\frac{1}{2}}$; $\varphi_0$, $\gamma_1$ and $\gamma_2$ are the parameters of $U$(see Lemma \ref{lemma2}); $k_1=1,2$ and $k_2=1,2$.
% with $\varphi_0$ as well as $\gamma_1$ and $\gamma_2$ is stated as in Lemma \ref{lemma2},
\end{thm}
The proof is in the Appendix \ref{pt1}.

It is easy to check that with different values of $k_1$ and $k_2$ of $V$, the obtained values of $C_{33\_V}(\ket{\phi})$ are the same. Due to this reason, we do not differentiate between the values of $k_1$ as well as $k_2$ in the following. Moreover, it is worth noting that, for any given $\mathcal{U}_k$, the number of satisfied $V$ is infinite since there are only two necessary restrictions imposed on the three free parameters of $V$ to promise $C_{33\_V}(U)=1$. And by Corollary \ref{c31}, the value of $C_{33\_V}(\ket{\phi})$ also does not depend on the respective $\psi$ or $\chi$ but on the sum of them(which holds also for quantum mixed states since quantum mixed states can be seen as a convex combination of pure quantum states). So it seems that we need not to care about the individual values of $\psi$ or $\chi$. However, we show that specifying a special value of $\chi$ or $\psi$ for $V$ can facilitate the discrimination task when $\bold{P}$ is initially in the plane $l(1)$ (see the discussion after Theorem \ref{th3}). The set of all the satisfied $V$ for $\mathcal{U}_k$ is denoted by $\mathcal{V}^k$. And the collection of $\mathcal{V}^k(k=1,2,3,4)$ is denoted by $\mathcal{V}$, i.e., $\mathcal{V}=\{\mathcal{V}^k|k=1, 2, 3, 4\}$.
%can divide all the above $V$ for $\mathcal{U}_k$ into four groups  denoted by $\mathcal{V}^k_i(i=1,2,3,4)$ in terms of the value of $k_1$ and $k_2$. Such division is helpful for the discrimination when $\bold{P}$ is initially in $l(1)$(see Theorem \ref{th3} below).
%All the above $V$ for $\mathcal{U}_k$ are divided into four groups in terms of the values of $k_1$ and $k_2$.
%denoted by $\mathcal{V}_i^k(i=1,2,3,4)$and $\forall V \in \mathcal{V}_i^k(i=1,2,3,4)$
\begin{thm}\label{th2}
Given two qubits in the state $\rho$, no unitary matrix $V\in \bold{U}(2)$ can make any entry of $\bold{P}_V(\rho)$ be 1, unless $\bold{P}(\rho)$ is in the plane $l(1)$.
\end{thm}
The proof is in the Appendix \ref{pt2}.

\begin{figure}
\centering
\includegraphics[width=3in,height=2in]{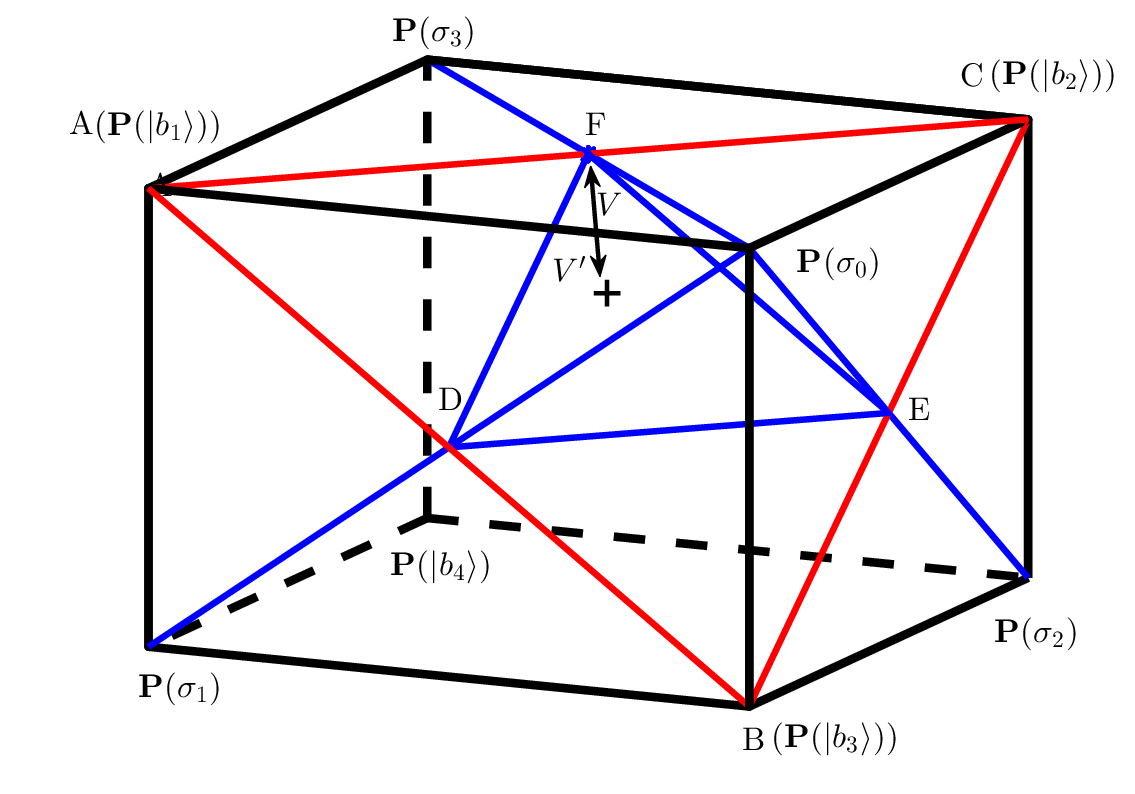}
\caption{The cases where $\bold{P}$ is initially in $l(1)$. Given the initial statistic $\bold{P}$(represented by the black symbol '+') of the two qubits in the plane $l(1)$, if there exists $V$ such that $\bold{P}_V$ is equal to the F point, i.e., $(0,0,1)'$ for both common cause $\rho$ and direct cause $U$ with the same value of $\bold{P}$, then with $V'$, $\bold{P}_V$ should be able to be transferred back to $\bold{P}$, which is a necessary condition for $\bold{P}_V(\rho) = (0,0,1)'$.}
\label{fig4}
\end{figure}

As a special case of Theorem \ref{th1}, when $\bold{P}$ is initially in the plane $l(1)$, $\bold{P}_V(U)$ is $(0,0,1)'$ by the obtained $V$. However, in this case, $\bold{P}_{V}(\rho)$ can also be $(0,0,1)'$, which may cause the discrimination task to fail. We discuss this special case in Theorem \ref{th3} and show the conditions under which the obtained $V$ can still work to promise $\bold{P}{_V}(\rho)$ not to be $(0,0,1)'$(see Fig. \ref{fig4}).
\begin{thm}\label{th3}
Given $\bold{P}\neq (0, 0, 1)$ in the plane $l(1)$, analyzing current $\bold{P}$ can obtain $\mathcal{U}_k(p_0, p_1, p_2, p_3)(k=1,2,3,4)$ and the corresponding $\mathcal{V}$ as stated above. For any quantum state $\rho$ satisfying $\bold{P}(\rho)=\bold{P}$ and $\forall V \in \mathcal{V}^k(k=1,2,3,4)$, $\bold{P}{_V}(\rho)\neq(0,0,1)'$ holds unless that $f_2=\sin(2\chi - 2\psi)=0$ or $f_2=f_3=0$, where $\psi$ and $\chi$ are parameters of $V$ as stated in Eq. \eqref{V}, $f_i(i=1,2,3)\in \mathbb{R}$ are parameters of $\rho_{_V}$ and
\begin{equation}\label{eqth32}
\rho_{_V} = \left(V \otimes V\right)'\rho \left(V \otimes V\right)=
\begin{pmatrix}
  f_1 & 0 & 0 & f_2 - if_3 \\
  0 & 0 & 0 & 0 \\
  0 & 0 & 0 & 0 \\
  f_2 + if_3 & 0 & 0 & 1-f_1
\end{pmatrix}.
\end{equation}
%Obviously, $\bold{P}(\Phi)=\bold{P}(\mathcal{U})$.
%$\sin(\omega_{\rho})=\frac{f_3}{\sqrt{f_2^2 + f_3^2}}$,
%$\cos(\omega_{\rho})=\frac{f_2}{\sqrt{f_2^2 + f_3^2}}$
%Further, under the condition $\sin(2\psi_1 + 2\chi_1)=0$, let $C_{22\_{V_1}}(\rho)=C_{22\_{V_1}}(U)$, we get $\sin^2(2\varphi)\cos^2(2\psi) + 2f_3\cos(2\varphi)\sin(4\psi)=\sin^2(2\varphi)\cos^2(2\psi)$ that
\end{thm}
The proof is in Appendix \ref{pt3}.

Following from Theorem \ref{th3}, there are three cases where $\bold{P}{_V}(\rho)$ can be equal to $(0,0,1)'$, including $\bold{P}=(0,0,1)'$, $f_2 = \sin(2\chi - 2\psi )=0$ and $f_2=f_3=0$, which becomes a barrier for the discrimination task.
For the first case, we can make $\bold{P}$ leave $(0,0,1)'$ by applying a proper unitary operation $V_0$ on the observables first, for example, $V_0$ with $\varphi=\pi/8$ and $\psi + \chi = 0$. Actually after having applied such $V_0$, the current $C_{33}$ becomes $\cos^2(\pi/4)$ regardless of what cause is actually behind the initial $\bold{P}$.

 For the second case, we can simply let $\sin(2\chi - 2\psi )\neq 0$, i.e., $\chi - \psi\neq \frac{k\pi}{2}$, where $k\in\mathbb{Z}$. Note there is no contradiction between this scheme with the design of unitary operators $V$ in Theorem \ref{th1}, because as we discussed earlier, the value of $\bold{P}_V$ depend not on the individual values of $\psi$ and $\chi$ but on the sum of them.

 For the third case, it is easy to check that with any $V_x\in \bold{U}(2)$, for any $U$ satisfying $\bold{P}(U)=(0,0,1)'$, $\bold{P}_{V_x}(\rho_{_V})$ is always equal to $\bold{P}_{V_x}(U)$. That is with $f_2=f_3=0$, no unitary operation $V$ can further diverge the measurement results of quantum common cause from the measurement results of quantum direct cause when their $\bold{P}$ are originally the same. Recall that the value of $\bold{P}_V$ is restricted to the same plane $l(b)$ when only applying a single $V$ on the observables of the two qubits respectively; that only in the plane $l(1)$, $\bold{P}_V(\rho)$ may be $(0,0,1)'$. Then a feasible solution to this case may be applying different unitary operators on the observables of one qubit and another qubit respectively to transfer current $\bold{P}$ to another plane $l(b)(b\neq 1)$. We choose the plane $l(-1)$ as the destination plane because in the plane, the corresponding destination point $(-1,-1,1)'$ is far from the image set of $\bold{P}_V(\rho)$, which may help to reduce the uncertainty caused by the quantum mechanism in the discrimination process. In addition, we only consider how to transfer $\bold{P}=(0,0,1)'$ to the plane $l(-1)$ since $\bold{P}$ can always be transferred to $(0,0,1)'$ first in this case and the analysis process is relatively simpler when compared with the cases that $\bold{P}$ is not $(0,0,1)'$. We have the following theorem.
\begin{thm}\label{th4}
Let $V_{_+}$ be an identity matrix and
\begin{equation}\label{th4v0}
V_{_-}=
\begin{pmatrix}
  0 & 1 \\
  1 & 0
\end{pmatrix}.
\end{equation}
Given two qubits either in the quantum state $\rho$ or existing a direct cause $U$ between them, where $\rho$ is stated as in Eq. \eqref{eqth32} with $f_2=f_3=0$ and $U$ satisfies $\bold{P}(U)=(0,0,1)'$, after having applied $V_{_-}$ and $V_{_+}$ on the observables of the two qubits in temporal order, then measuring these new observables gives new values of $\bold{P}$, which are in the plane $l(-1)$.
\end{thm}
The proof is in the Appendix \ref{pt4}.

Once $\bold{P}$ is transferred to the plane $l(-1)$, we can conveniently use Theorem \ref{th1} and Theorem \ref{th2} to distinguish quantum direct causes from quantum common causes.
%The measure of the set of $\bold{P}$ in this case make up a measure$the 16 groups of unitary operators $\mathcal{V}^j_i$($i,j\in \{1,2,3,4\}$) can be obtained. If there is a direct quantum causal connection $U \in \mathcal{U}_k$ between the two qubits, then for $\forall i\in\{1,2,3,4\}$ and $\forall V \in \mathcal{V}^k_i$, $\bold{P}$ can be transferred to the point F, i.e., $(0,0,1)'$. Otherwise, If the two qubits have a common cause, then there exists $m\in\{1,2,3,4\}$. With $\forall V \in \mathcal{V}^k_m$, $\bold{P}_V$ (represented by the red symbol '.') is not equal to the point F. This conclusion does not hold when $\bold{P}$ is initially in the point F or in the line segment DE.
\section{Discrimination Method}
Based on the above theoretical observations, we develop a method for the discrimination task by using experimental observations only.
Supposing we have prepared many copies of a system to be tested, we measure the same Pauli observables on the two qubits before unitary transformation to get the estimated values of $\bold{P}$
and measure the transformed Pauli observables on the two qubits to get the estimated values of $\bold{P}_V$ if it is necessary. With the estimated values of $\bold{P}$ and $\bold{P}_V$, we identify whether there are causalities between the two qubits or not. It contains the following steps(see Fig. \ref{Fig5}):

(1) Measure the same Pauli observable $\sigma_i(i=1,2,3)$ on the two qubits to get the estimated value of $\bold{P}$.

(2) If the estimated value of $\bold{P}$ is outside of the overlapping area, it is the end; if the estimated value of $\bold{P}$ is $(0,0,1)'$, then apply a unitary operation $V_0$ as stated in the discussion after Theorem \ref{th3} on the observables of the two qubits first to get new observables and new $\bold{P}$ whose third entry is no longer 1; else, go to the next step.

(3) Using Eq. \eqref{Uform} to obtain the set of possible cases of quantum causality, i.e., $\mathcal{U}(p_0,p_1,p_2, p_3)$.

(4) Following from Corollary \ref{c31}, divide $\mathcal{U}$ into four subsets $\mathcal{U}_k(k=1,2,3,4)$. For every $\mathcal{U}_k(k=1,2,3,4)$, by Eq. \eqref{th11} in Theorem \ref{th1}, design one group of unitary matrices $\mathcal{V}^k$.

(5) $\forall k\in\{1,2,3,4\}$, pick one $V^k\in \mathcal{V}^k$ with $0 < \chi - \psi < \frac{\pi}{2}$ at random; apply it on the current observables to get the estimated value of $\bold{P}_V$ denoted by $\bold{P}^k$.

(6) If there exists a $k\in\{1,2,3,4\}$ such that $\bold{P}^k=(0,0,1)'$, then after having applied $V^k$ on the current two observables, apply $V_{_-}$ and $V_{_+}$ as stated in Theorem \ref{th4} on the current observables to get new observables and new value of $\bold{P}$; go to the next step. Otherwise, if there exists a $k\in\{1,2,3,4\}$ such that the third entry of $\bold{P}^k$ is 1 and its first two entries are equal, then there is a direct causal connection between the two qubits; else there is a common cause acting on them.

(7) For the current observables and the current value of $\bold{P}$, perform steps (3) through (5) to get the new value of $\bold{P}^k$. If there exists a $k\in\{1,2,3,4\}$ such that the third entry of $\bold{P}^k$ is 1, then there is a direct causal connection between the two qubits; else there is a common cause acting on them.

Simulation experiments were conducted on the systems with the parameters of the quantum states (common causes) and the unitary matrices (direct causes) randomly sampled from their legal intervals. For each measurement, we simulated it by sampling 200 examples from $\bold{P}$ or $\bold{P}_V$ (which actually includes three distributions) thereby getting the estimated value of $\bold{P}$ or $\bold{P}_V$. In total, we created 10000 quantum states and 10000 unitary matrices respectively. The tolerance of the algorithm was set as 0.1, which means that if $|a-b|<0.1, a, b\in \mathbb R$, we argue $a=b$. Consequently, given two vectors $\bold{P}_{a}=(a_1,a_2,a_3)'$ and $\bold{P}_{b}=(b_1, b_2, b_3)'$, they were considered to be equal, if $\forall i\in\{1,2,3\} $, $|a_i-b_i|<0.1$. The relatively loose tolerance can prevent a bad immediate discrimination conclusion when the estimated $\bold{P}$ is near $(0,0,1)'$; and $\bold{P}$ in these cases would be transferred to the planes near $l(-1)$ [see step (6)], where a reliable discrimination can always be obtained even the tolerance is relatively big. Each experiment was repeated five times. The average number of failed cases is $251(\pm10)$, accounting for $1.26\%(\pm0.05\%)$. And when the number of sampling increased to more than 800, no failure cases were observed.

% Define block styles
\tikzstyle{decision} = [diamond, draw, fill=blue!20, text width=4.5em, text badly centered, node distance=3cm, inner sep=0pt]
\tikzstyle{block} = [rectangle, draw, fill=red!20, text width=8em, text centered, minimum height=1.5em]
\tikzstyle{line} = [draw, -latex']
\tikzstyle{elli} = [draw, ellipse,fill=blue!20, node distance=3cm, minimum height=2em]
\tikzstyle{io} = [trapezium, trapezium left angle=70, trapezium right angle=110, minimum width=1.5cm,  text centered, draw=black, fill=blue!30]
{
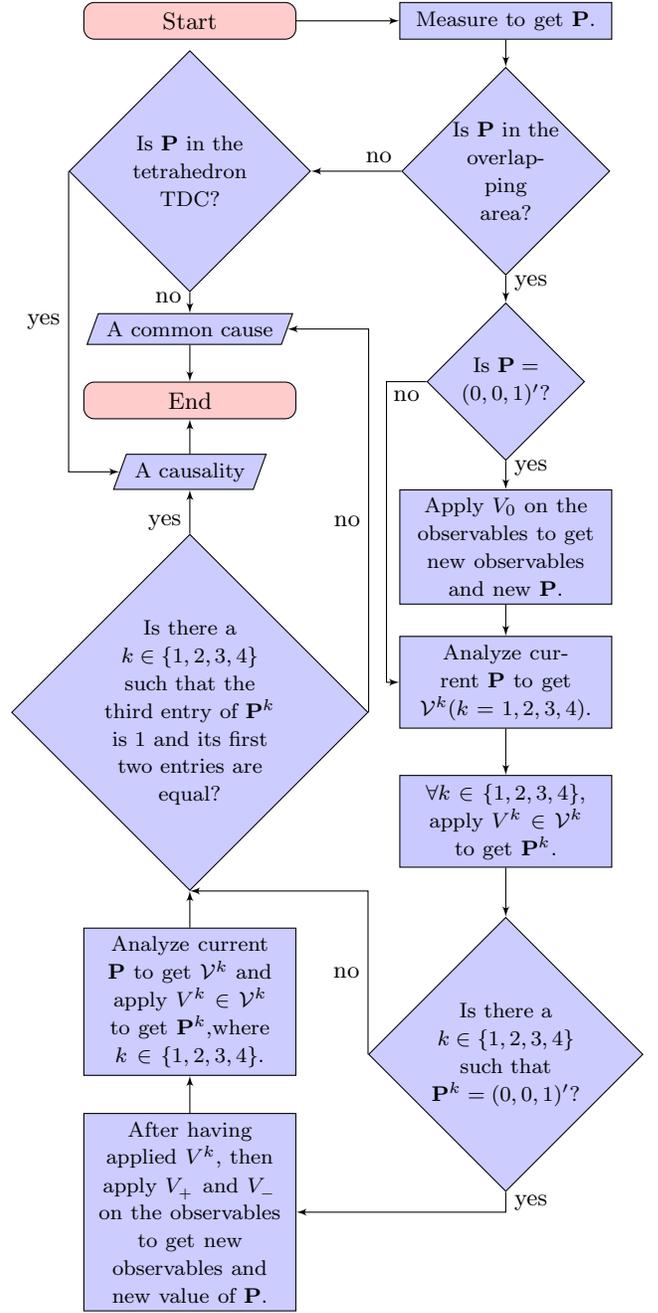
\begin{figure}
\centering
\begin{tikzpicture}[node distance = 1.5cm, auto]
    % Place nodes
    \node [block,  rounded corners] (start) {Start};
    \node [block, fill=blue!20, right of=start, node distance=4.2cm] (measurep) {\footnotesize{Measure to get $\bold{P}$.}};
    \node [decision, fill=blue!20,node distance=2cm,below of=measurep] (isoverlap) {\footnotesize{Is $\bold{P}$ in the overlapping area?}};
    \node [decision, fill=blue!20,node distance=2.8cm, below of=isoverlap] (third) {\footnotesize{Is $\bold{P}=(0,0,1)'$?}};

    \node [block, fill=blue!20,below of=third, node distance=2.2cm] (V0) {\footnotesize{Apply $V_0$ on the observables to get new observables and new $\bold{P}$.}};
    \node [block, fill=blue!20, text width=8em,node distance=1.8cm, below of=V0] (getU) {\footnotesize{Analyze current $\bold{P}$ to get $\mathcal{V}^k(k=1,2,3,4)$.}};
    \node [block, fill=blue!21, node distance=1.85cm, below of=getU] (remeasure) {\footnotesize{$\forall k\in\{1,2,3,4\}$, apply $V^k\in\mathcal{V}^k$ to get $\bold{P}^k$.}};

    \node [decision,fill=blue!20, node distance=2cm,text width=7em,below of=start] (isinTDC) {\footnotesize{Is $\bold{P}$ in the tetrahedron TDC?}};

    \node [io, fill=blue!20,below of=isinTDC, node distance=2.1cm] (cc) {\footnotesize{A common cause}};
    \node [block,  rounded corners, below of=cc, node distance=0.95cm] (end) {End};
    \node [io, fill=blue!20,below of=end, node distance=0.95cm] (dc) {\footnotesize{A causality}};
    \node [decision,fill=blue!20, node distance=3.2cm,text width=7em,below of=dc] (iscausality) {\footnotesize{Is there a $k\in\{1,2,3,4\}$ such that the third entry of $\bold{P}^k$ is 1 and its first two entries are equal?}};
    %\node [decision,fill=blue!20, node distance=2.8cm,text width=7em,below of=dc] (iscausality) {\footnotesize{Is there $k\in\{1,2,3,4\}$ such that the third entry of $\bold{P}^k=(p,p,1)(-1\leq p\leq 1)$ }};

    \node [decision,fill=blue!20, node distance=3.1cm,text width=7em,below of=remeasure] (is001) {\footnotesize{Is there a $k\in\{1,2,3,4\}$ such that $\bold{P}^k=(0,0,1)'$?}};
    \node [block, fill=blue!20, text width=8em,node distance=3.85cm, below of=iscausality](perform35){\footnotesize{Analyze current $\bold{P}$ to get $\mathcal{V}^k$ and apply $V^k\in\mathcal{V}^k$ to get $\bold{P}^k$,where $k\in\{1,2,3,4\}$.}};
    \node [block, fill=blue!20, node distance=2.8cm, below of=perform35] (applyV+-) {\footnotesize{After having applied $V^k$, then apply $V_{_+}$ and $V_{_-}$ on the observables to get new observables and new value of $\bold{P}$.}};

%    \node [elli, left of=start] (expert) {expert};
%    \node [elli, right of=start] (system) {system};
%    \node [block, below of=start] (identify) {identify candidate models};
%    \node [block, below of=identify] (evaluate) {evaluate candidate models};
%    \node [block, left of=evaluate, node distance=3cm] (update) {update model};
%    \node [decision, below of=evaluate] (decide) {is best candidate better?};
%    \node [block, below of=decide, node distance=3cm] (stop) {stop};
%    % Draw edges
     \path [line] (start) -- (measurep);
     \path [line] (measurep) -- (isoverlap);
     \path [line] (isoverlap) -- node[near start, anchor = west] {yes} (third);
     \path [line] (third) -- node[near start] {yes} (V0);
     \path [line] (third.west)-| node[near start, anchor=north] {no}($(getU.west) + (-0.17,0)$)--(getU.west);
     \path [line] (V0) -- (getU);
     \path [line] (getU) -- (remeasure);
     %\path [line] (remeasure.west) -| (iscausality.south);
     \path [line] (remeasure.south) -- (is001.north);  %add
     \path [line] (is001.south)|-node[near start]{yes}(applyV+-.east);  %add
     \path [line] (applyV+-.north)--(perform35.south);  %add
     \path [line] (perform35.north)--(iscausality.south);  %add
     \path [line] (is001.west)|-node[near start]{no}(iscausality.south);  %add

     \path [line] (isoverlap) --node[near start, anchor=south] {no} (isinTDC);
     \path [line] (iscausality.north) --node[near start, anchor=east] {yes} (dc.south);
     \path [line] (iscausality.east) |-node[near start, anchor=east] {no} (cc.east);
     \path [line] (isinTDC.south) --node[near start, anchor=east] {no} (cc.north);
     \path [line] (isinTDC.west) |-node[near start, anchor=east] {yes} (dc.west);
     \path [line] (cc.south) -- (end.north);
     \path [line] (dc.north) -- (end.south);
     %\path [line] (isinTDC) |-node[near start, anchor=east] {yes} (dc);([xshift=1cm]p2.south)
%    \path [line] (identify) -- (evaluate);
%    \path [line] (evaluate) -- (decide);
%    \path [line] (decide) -| node [near start] {yes} (update);
%    \path [line] (update) |- (identify);
%    \path [line] (decide) -- node {no}(stop);
%    \path [line,dashed] (expert) -- (start);
%    \path [line,dashed] (system) -- (start);
%    \path [line,dashed] (system) |- (evaluate);
\end{tikzpicture}
\caption{The flow chart of the proposed method}\label{Fig5}
\end{figure}}

% the only way to distinguish between the
%two possibilities is to replace observation of the early variable with
%an intervention on it
%In quantum cases, Bell theorem rules out the classical
%common cause explanation of the correlations that vio-
%late a bell inequality. To make causal model be compat-
%ible with quantum mechanism, a considerable e

%been recently devoted to transfer classical causal mod-
%eling technique to the quantum domain, including the
%introducing of hidden, ne tuned mechanisms and refor-
%mulating causal models from the ground up by direct use
%of the quantum formalism
%
%any product state is a non-commutative common
%cause for some given set of correlations; secondly, any product state is a such
%a cause for any correlations obtained by measurements on a bipartite quantum
%state. Furthermore, the notion that the common causes can in any way causally
%reproduce or explain the correlations is lost in this approach
%
%We exploit these facts to
%devise, for a particular class of causal scenarios, a complete solution
%of the causal inference problem using observation alonea task
%that is impossible classically. We implement a family of such causal
%scenarios experimentally and confirm our conclusion.
%\section*{References}
\section{Conclusions and Future Works}

%Finding causal explanations of observed correlations is an important problem across a broad range of disciplines. In classical causal models,
The possibility of intervening is requisite for causal reasoning of classical causal models. However, the interventionist schemes cannot be directly applied to the quantum case. The dilemma is presented as a choice between relinquishing one of two assumptions: the causal Markov condition or faithfulness (no-fine-tuning) \citep{shrapnel2015discovering}. %It is not possible to explain Bell correlations with classical causal models without unwelcome fine-tuning of the parameters.
Instead of trying to modify one of the existing assumptions, another probably better approach to avoid such a dilemma is reformulating causal models in a way that makes direct use of the quantum formalism and providing a quantum interventionist framework for Bayesian inference as well as causal inference \citep{pienaar2017causality}.
%The two types of inference, connected by Reichenbach's principle, are the heart of causal models.
%The aim of them is to provide a quantum framework for Bayesian inference and causal inference, which are the heart of causal models.
%interventionist account of causal explanation fails to explain quantum phenomena
%similar causal inferring models are developed. These quantum interventionist way extends the classical interventionist methods by introducing hidden and fine tuned mechanisms or developing the quantum versions of causal models\citep{costa2016quantum}. However, interventionist schemes are often difficulty to be implemented due to the technical or ethical reasons.

In this paper, distinct from the quantum interventionist framework, we adopt the frequentist manner and prove that quantum observational schemes can universally distinguish causality from correlations. We first analyze the way in which the statistic $\bold{P}$ moves when the observables are transformed by unitary operations. Using this obtained property, we show how to design unitary matrices to make quantum common causes and quantum causality be distinguishable. A general method is developed to distinguish the two and is testified by simulation experiments. Nonetheless, the mixture case of quantum common causes and quantum direct causes may also account for the observed correlation, which was not discussed in this paper. We leave its analysis and the method development in the future work.
%This property provides a basis for designing the unitary matrix $V$. Specially, in the plane $l(1)$.%in theory and if taking possible measurement errors into account, we conclude they are separable asymptotically.
\begin{acknowledgments}
This work is funded in part by the National Key R\&D Program of China(2017YFE0111900), the National Natural Science Foundation of China(61876129, 61650303), the National Natural Science Foundation of China(Key Program, U1636203), the Alibaba Innovation Research Foundation 2017 and the European Unions Horizon 2020 research and innovation programme under the Marie Skodowska-Curie grant agreement No. 721321. Part of the work was performed when Yuexian Hou and Chenguang Zhang visited the Open University during June-July 2018. We also thank the anonymous reviewers for their insightful comments and suggestions.
\end{acknowledgments}

\appendix

\section{\label{pl1}PROOF OF LEMMA 1}
\begin{proof}
On the one hand, a straightforward calculation can show
\begin{equation}
\label{ZEqnNum492291}
\bold{P}\left({\left| \phi  \right\rangle} \right)=\sum _{i=1}^{4}w_{i}^{2}\bold{P}\left({\left|
b_{j} \right\rangle} \right)
\end{equation}
holds for $\phi =\sum _{j=1}^{4} w_{j} e^{i\theta _{j} } {\left| b_{j}  \right\rangle} $.
Then the existence is proven. On the other hand, $\left\{\bold{P}\left({\left| b_{i}  \right
\rangle} \right)|i=1,2,3,4\right\}$ form a complete basis; if there exists another
group of coefficients $\left\{v_{j} |j=1,2,3,4,v_{j} \in {\rm {\mathcal C}}\right
\}$ such that ${\left| \phi  \right\rangle} =\sum _{j=1}^{4} v_{j} {\left| b_{j}
\right\rangle} $ satisfies Eq. \eqref{ZEqnNum492291}, thus $|v_{j} |^{2} =w_{j}^{2} $.
Then the uniqueness is proven\textit{.}
\end{proof}

\section{\label{pl2}PROOF OF LEMMA 2}
\begin{proof}
 Let \begin{equation} \label{2)}
U_x=\begin{pmatrix}
  a_1+ia_2 & b_1+ib_2\\
  -e^{i\alpha}(b_1 - ib_2) & e^{i\alpha}(a_1 - ia_2)
\end{pmatrix}
\end{equation}
be an arbitrary unitary matrix in $\bold{U}(2)$, where $a_{1}^{2} +a_{2}^{2}
+b_{1}^{2} +b_{2}^{2} =1$. As proven in Ref. \cite{hu2018discrimination},
\begin{equation} \label{ZEqnNum109774}
\bold{P}(U_x)=\begin{pmatrix}
  2(c-d) - 1\\
  2(c+d)+1\\
  2(a_1^2 + a_2^2)-1
\end{pmatrix}
\end{equation}
where $c=\frac{1}{2} +a_{1} a_{2} \sin \alpha +\frac{\cos \alpha }{2} (a_{1}^{2} -a_{2}^{2}
)$,$d=b_{1} b_{2} \sin \alpha +\frac{\cos \alpha }{2} (b_{1}^{2} -b_{2}^{2} )$. Plug $U$ defined in Eq. \eqref{Uform} into above Eq. \eqref{ZEqnNum109774}, it is easy to find
\begin{equation}
\label{ZEqnNum529269}
\bold{P}(U)=\sum _{j=0}^{3}p_{j} \bold{P}(\sigma _{j} )
\end{equation}
holds, i.e., the existence of $U$ is proven, where $\bold{P}(\sigma _{0} )=(1,1,1)'$,$\bold{P}(
\sigma _{1} )=(1,-1,-1)'$, $\bold{P}(\sigma _{2} )=(-1,1,-1)'$, $\bold{P}(\sigma _{3} )=(-1,-1,1)'$.
Next to prove the uniqueness of $U$. Supposing an unknown $U_{x} $ satisfies above
Eq. \eqref{ZEqnNum529269}, then
\begin{equation} \label{ZEqnNum530066}
 \begin{cases}
    a_1a_2\sin\alpha + (a_1^2 - a_2^2)= p_0 - p_3 \\
    a_1^2 + a_2^2=p_0+p_3\\
    b_1b_2\sin\alpha + (b_1^2 - b_2^2)= p_2 - p_1.\\
    b_1^2 + b_2^2=p_1+p_2
 \end{cases}
\end{equation}
If $p_{0} +p_{3} =0$ or $p_{1} +p_{2} =0$, we have $a_{1} =a_{2} =0$ or $b_{1} =b_{2}
=0$. Otherwise, let $c_{1} =p_{0} +p_{3} $, $c_{2} =p_{0} -p_{3} $, $d_{1} =p_{1}
+p_{2} $, $d_{2} =p_{2} -p_{1} $, $\cos (\zeta _{1} )=\frac{a_{1} }{\sqrt{c_{1} }
} $,$\cos (\zeta _{2} )=\frac{b_{1} }{\sqrt{d_{1} } } $,$\sin (\zeta _{1} )=\frac{a_{2}
}{\sqrt{c_{1} } } $ and $\sin (\zeta _{2} )=\frac{b_{2} }{\sqrt{d_{1} } } $; plug
these equations into above Eq. \eqref{ZEqnNum530066} and assume $\alpha =\theta $ is
known, thus $\zeta _{1} =\gamma _{1} +\frac{\theta }{2} $, $\zeta _{2} =\gamma _{2}
+\frac{\theta }{2} $, where $\gamma _{1} =\frac{(-1)^{n_{1} } \arccos \left(\frac{c_{2}
}{c_{1} } \right)}{2} +k_{1} \pi $, $\gamma _{2} =\frac{(-1)^{n_{2} } \arccos \left(
\frac{d_{2} }{d_{1} } \right)}{2} +k_{2} \pi $ and $n_{1} ,n_{2} ,k_{1} ,k_{2} \in
\{ 0,1\}$.
\end{proof}

\section{\label{pl3}PROOF OF LEMMA 3}
\begin{proof}
After having applied unitary evolution $V\otimes V$ on
the observable $\sigma _{i} \otimes \sigma _{i} $, the probability of finding both
qubits in the same direction is
%\begin{subequations}
\begin{eqnarray}
&&p_{_{V} } (k=m|ii)={\rm Tr}\left(\left(V\otimes V\right)\left(
\xi _{i}^{dd} +\xi _{i}^{uu} \right)\left(V\otimes V\right)^{{'} } \rho \right)
\nonumber\\
&&={\rm Tr}\left(\left(V\otimes V\right)\left(\xi _{i}^{dd} +\xi _{i}^{uu} \right)
\left(V\otimes V\right)^{{'} } {\left| \phi  \right\rangle} {\left\langle \phi  \right|}
\right)
\nonumber\\
&&={\rm Tr}\left(B'\left(V\otimes V\right)\left(\xi _{i}^{dd} +\xi _{i}^{uu}
\right)\left(V\otimes V\right)^{{'} } Bww'\right)\label{6)}.
\end{eqnarray}
%\end{subequations}
\end{proof}

\section{\label{pc32}PROOF OF COROLLARY 3.2}
\begin{proof}
We only need to prove that the sum of the three entries
of $\bold{P}_{V} \left({\left| \phi  \right\rangle} \right)$ is a fixed value independent
of $V$as well as the phase of $w_{j} (j=1,2,3,4)$.
\begin{equation} \label{ZEqnNum757428}
\resizebox{\linewidth}{!}{$
\begin{array}{ll} & {C_{11\_ V} \left({\left| \phi  \right\rangle} \right)+C_{22
\_ V} \left({\left| \phi  \right\rangle} \right)+C_{33\_ V} \left({\left| \phi  \right
\rangle} \right)} \\ {} & {=2\left(\sum _{i=1}^{3} p_{_{V} } (k=m|ii)\right)-3} \\
{} & {=2{\rm Tr}\left(B^{{\rm T}} \left(V\otimes V\right)\sum _{i=1}^{3} \left(\xi
_{i}^{dd} +\xi _{i}^{uu} \right)\left(V\otimes V\right)^{{'} } Bww'\right)-3.} \end{array}$}
\end{equation}
After
careful calculation, it is easy to find
\begin{equation} \label{ZEqnNum781437}
\resizebox{\linewidth}{!}{$
\begin{array}{l} {B^{{\rm T}} \left(V\otimes V\right)\sum _{i=1}^{3} \left(\xi _{i}^{dd}
+\xi _{i}^{uu} \right)\left(V\otimes V\right)^{{'} } B=\left(\begin{array}{cccc}
{2} & {0} & {0} & {0} \\ {0} & {2} & {0} & {0} \\ {0} & {0} & {2} & {0} \\ {0} &
{0} & {0} & {0} \end{array}\right).} \end{array}$}
\end{equation}

\noindent Plug Eq. \eqref{ZEqnNum781437} into Eq. \eqref{ZEqnNum757428}, we have
\begin{equation} \label{9)}
\begin{array}{ll} {} & {C_{11\_ V} \left({\left| \phi  \right\rangle} \right)+C_{22
\_ V} \left({\left| \phi  \right\rangle} \right)+C_{33\_ V} \left({\left| \phi  \right
\rangle} \right)} \\ {} & {=4\sum _{i=1}^{3} w_{i}^{2} -3=1-4w_{4}^{2} .} \end{array}
\end{equation}
\end{proof}

\section{\label{pl4}PROOF OF LEMMA 4}
\begin{proof}
It has been shown in Ref. \cite{hu2018discrimination} that $\bold{P}_{V} (U)=\bold{P}(V'UV)$. Further,
let $V'UV$ be $\left(\begin{array}{cc} {a_{1} +ia_{2} } & {b_{1} +ib_{2} } \\ {b_{1}
-ib_{2} } & {a_{1} -ia_{2} } \end{array}\right)$, then $a_{1} $, $a_{2} $, $b_{1} $ and $b_{2} $ would
take values as given in the Lemma \ref{lemma4}. Finally, by Eq. \eqref{ZEqnNum109774}, $C_{ii
\_ V} (U)(i=1,2,3)$, i.e., $C_{ii} (V'UV)(i=1,2,3)$ can be expressed as the function
of $a_{1} $, $a_{2} $, $b_{1} $ and $b_{2} $ as stated in Lemma \ref{lemma4}.
\end{proof}

\section{\label{pc41}PROOF OF COROLLARY 4.1}
\begin{proof}
From Lemma \ref{lemma2}, we see that ${\mathcal U}(p_{0} ,p_{1}
,p_{2} ,p_{3} )$ contains 16 unitary matrices(whose global phases are omitted), each
of which is determined by $(n_{1} ,n_{2} ,k_{1} ,k_{2} )$. Further, it is easy to
see that the unitary matrix determined by $(1-n_{1} ,n_{2} ,1-k_{1} ,k_{2} )$ is
the same as the unitary matrix determined by $(n_{1} ,n_{2} ,k_{1} ,k_{2} )$; and
the unitary matrix determined by $(n_{1} ,n_{2} ,1-k_{1} ,1-k_{2} )$ differs from
that determined by $(n_{1} ,n_{2} ,k_{1} ,k_{2} )$ only by the sign. Moreover, following
from Lemma 4, $\bold{P}_{V} (-U)=\bold{P}_{V} (U)$ holds. Thus, ${\mathcal U}(p_{0} ,p_{1}
,p_{2} ,p_{3} )$ can be divided into four subsets. Over each subset, the values of $\bold{P}_{V}
(U)$ are the same.
\end{proof}

\section{\label{pc42}PROOF OF COROLLARY 4.2}
\begin{proof}
We only need to prove that the sum of the three entries
of $\bold{P}_{V} (U)$ is a fixed value independent of $V$. Following from Lemma \ref{lemma4} and by
the fact that $a_{1}^{2} +a_{2}^{2} +b_{1}^{2} +b_{2}^{2} =1$, we get $C_{11\_ V}
(U)+C_{22\_ V} (U)+C_{33\_ V} (U)=4a_{1}^{2} -1$. Finally, by Lemma \ref{lemma2}, we have $a_{1}^{2}
=\cos (\varphi _{0} )\cos (\gamma _{1} )=p_{0} $ is dependent of the choice of $V$.
\end{proof}

\section{\label{pl5}PROOF OF LEMMA 5}
\begin{proof}
Corollary \ref{c32} and Corollary \ref{c42} have shown the normal vectors
of the discussed planes in the two cases are all $(1,1,1)'$, thus what we need to
prove is the constant terms of the two planes are actually equal. In fact, because $\Phi(w_{1} ,w_{2} ,w_{3} ,w_{4} )$ and ${\mathcal U}(p_{0} ,p_{1} ,p_{2} ,p_{3}
)$ are the analysis results of the same value of $\bold{P}$, then for any ${\left| \phi
\right\rangle} \in \Phi (w_{1} ,w_{2} ,w_{3} ,w_{4} )$ and for any $U\in {\mathcal U}(p_{0} ,p_{1} ,p_{2} ,p_{3} )$, we have $\bold{P}\left({\left| \phi\right\rangle} \right)=\bold{P}(U)$
thereby getting $e\bold{P}\left({\left| \phi \right\rangle} \right)=e\bold{P}(U)$,
where $e=(1,1,1)$. By Corollary \ref{c32} and Corollary \ref{c42}, we obtain $4p_{0} -1=1-4w_{4}^{2} $.
That's to say that the two planes have the same constant term. Further, $p_{0} $ and $w_{4}^{2} $ are
all nonnegative, so the range of the constant term must be $[-1,1]$.
\end{proof}

\section{\label{pl6}PROOF OF LEMMA 6}
\begin{proof}
According to the Lemma \ref{lemma5}, we only need to prove it holds for mixed quantum states. In fact, When $\rho$ is a mixed state, $\bold{P}(\rho )$ and $\bold{P}_{V}
(\rho)$ can be respectively regarded as a convex combination of $\bold{P}(\rho _{i})(i=1,2,...,N)$ and
a convex combination of $\bold{P}_{V} (\rho _{i} )(i=1,2,...,N)$ with the same combinatorial coefficients,
where $\rho _{i} (i=1,2,...,N)$ is a pure state and $N$ is the number of pure states.
Because by Lemma \ref{lemma5}, $\bold{P}_{V} (\rho _{i} )$ and $\bold{P}(\rho _{i} )$ are all in the same
plane $l(b_{i})$ (supposing $\bold{P}(\rho _{i} )$ is initially in the plane $l(b_{i} )$),
then the respective combinations of them with the same combinatorial coefficients should be in the same plane $l(b)$.
\end{proof}

\section{\label{pt1}PROOF OF THEOREM 1}
\begin{proof}
Recall that $\bold{P}_{V} (U)=\bold{P}(V'UV)$ should lie in the regular
tetrahedron(denoted by TDC) with vertices $\bold{P}(\sigma _{0} )=(1,1,1)'$,$\bold{P}(\sigma _{1}
)=(1,-1,-1)'$, $\bold{P}(\sigma _{2} )=(-1,1,-1)'$ and $\bold{P}(\sigma _{3} )=(-1,-1,1)'$. Also,
According to Corollary \ref{c42}, $\bold{P}_{V} (U)$ should be in the plane $l(4p_{0} -1)$. The
intersection of them is a triangle with vertices $(1,2p_{0} -1,2p_{0} -1)'$, $(2p_{0}
-1,1,2p_{0} -1)'$ and $(2p_{0} -1,2p_{0} -1,1)'$(see Fig. \ref{figf2}). So
if $C_{33} (V'UV)=1$, the other two entries must be $2p_{0} -1$.

Next, we prove there exists $V$ as stated in Eq. \eqref{V} such that $C_{33\_ V} (U)=1$, where $U\in {\mathcal U}_{k}$ is
given as in Eq. \eqref{Uform}. It has been presented in Lemma \ref{lemma4} that $C_{33\_ V} (U)=2(a_{1}^{2} +a_{2}^{2} )-1$, where $a_{1} =\cos (\varphi _{0}
)\cos (\gamma _{1} )$ is independent of $V$. And for $a_{2} $, we have
\begin{equation} \label{10)}
\begin{array}{ll}
a_{2}=\cos (\varphi _{0} )\cos (2\varphi )\sin (\gamma _{1} )\\
-\sin (\varphi _{0})\sin (2\varphi )\sin (\gamma _{2} -\psi -\chi )=r\cos (2\varphi +\omega ),
\end{array}
\end{equation}
where
\begin{subequations}
\begin{equation} \label{11)}
r=\sqrt{\mathop{\cos }\nolimits^{2} (\varphi _{0} )\mathop{\sin }\nolimits^{2} (
\gamma _{1} )+\mathop{\sin }\nolimits^{2} (\varphi _{0} )\mathop{\sin }\nolimits^{2}
(\gamma _{2} -\psi -\chi )} ,
\end{equation}
\begin{equation} \label{12)}
\sin (\omega )=\frac{\sin (\varphi _{0} )\sin (\gamma _{2} -\psi -\chi )}{r} ,
\end{equation}
\begin{equation}
\label{13)}
\cos (\omega )=\frac{\cos (\varphi _{0} )\sin (\gamma _{1} )}{r} .
\end{equation}
\end{subequations}
Because $|\cos (2\varphi +\omega )|\le 1$, we must promise $r^{2} \ge 1-a_{1}^{2} $ to
get $a_{1}^{2} +a_{2}^{2} =1$. Simplifying $r^{2} \ge 1-a_{1}^{2} $, we get its equivalent
form
\begin{equation} \label{ZEqnNum743199}
\mathop{\sin }\nolimits^{2} (\gamma _{2} -\psi -\chi )\ge 1.
\end{equation}
That's to say $r^{2} \ge 1-a_{1}^{2} $ is possible only when
\begin{equation} \label{ZEqnNum247707}
\psi +\chi =\gamma _{2} -\frac{k_{1} \pi }{2} ,
\end{equation}
where $k_{1} =1,3$. At this moment, $1-a_{1}^{2} $ is in fact the maximum value of $r^{2} $.
Consequently, it demands $|\cos (2\varphi +\omega )|=1$, i.e.,
\begin{equation} \label{ZEqnNum160307}
\varphi =\frac{k_{2} \pi -\omega }{2} ,
\end{equation}
where $k_{2} =0,1$. Taken together, the legal $V$ should meet the Eq. \eqref{ZEqnNum247707} and
Eq. \eqref{ZEqnNum160307} simultaneously.
\end{proof}

\section{\label{pt2}PROOF OF THEOREM 2}
\begin{proof}
If $\rho $ is a pure quantum state, denote it by ${\left|\phi  \right\rangle} $ and suppose ${\left| \phi  \right\rangle} \in \Phi (w_{1},w_{2} ,w_{3} ,w_{4} )$, where $w_{j} \in  {\mathbb R}(j=1,2,3,4)$
and $\sum_{j=1}^{4} w_{j}^{2} =1$. Recall that $\bold{P}_{V} \left({\left| \phi  \right\rangle} \right)=\bold{P}
\left(\left(V\otimes V\right)^{{'} } {\left| \phi  \right\rangle} \right)$ should
lie in the regular tetrahedron(denoted by TCC) with vertices $\bold{P}\left({\left| b_{1}
\right\rangle} \right)=(1,-1,1)'$,$\bold{P}\left({\left| b_{2}  \right\rangle} \right)=(-1,1,1)'$, $\bold{P}
\left({\left| b_{3}  \right\rangle} \right)=(1,1,-1)'$ and $\bold{P}\left({\left| b_{4}\right\rangle} \right)=(-1,-1,-1)'$. Meanwhile,
as presented in Corollary \ref{c32}, $\bold{P}_{V}\left({\left| \phi  \right\rangle} \right)$ should also be in the plane $l(1-4w_{4}^{2})$.
The intersection of TCC and $l(1-4w_{4}^{2} )$ is a triangle with vertices $(-1,1-2w_{4}^{2},1-2w_{4}^{2} )^{'} $, $(1-2w_{4}^{2} ,-1,1-2w_{4}^{2} )^{'} $ and $(1-2w_{4}^{2},1-2w_{4}^{2} ,-1)^{'}$ (see Fig. \ref{figf2}). Obviously, any linear combination
of the three vertices can not be a vector with any entry being 1 ,unless $w_{4} =0$,
i.e., unless $\bold{P}\left({\left| \phi  \right\rangle} \right)$ is in the plane $l(1)$.

If $\rho $ is a mixed quantum state, it is easy to check $\bold{P}_{V}(\rho)$ is a convex combination
of $\bold{P}_{V}(\rho_{i})(i=1,2,...,N)$, where $\rho _{i}$ is a pure quantum state
and $N$ is the number of pure quantum states. Since except the case that $\bold{P}(\rho
_{i} )$ is in the plane $l(1)$, any entry of $\bold{P}_{V} (\rho _{i} )$ is
not 1, we have any entry of $\bold{P}_{V} (\rho )$ that is the convex combination of $\bold{P}_{V}
(\rho _{i} )$ should not be 1, unless $\bold{P}(\rho )$ is in the plane $l(1)$.
\end{proof}

\section{\label{pt3}PROOF OF THEOREM 3}
\begin{proof}
We first prove that if $\bold{P}(\rho _{_{V} } )=(0,0,1)'$, $\rho_{_{V} } =\left(V\otimes V\right)^{{'} } \rho \left(V\otimes V\right)$ can be expressed
as \begin{equation} \label{ZEqnNum494785}
\left(\begin{array}{cccc} {f_{1} } & {0} & {0} & {f_{2} -if_{3} } \\ {0} & {0} &
{0} & {0} \\ {0} & {0} & {0} & {0} \\
{f_{2} +if_{3} } & {0} & {0} & {1-f_{1} } \end{array}
\right).
\end{equation}
Here, $\rho _{_{V} } $ may be a mixed quantum state. And suppose it is a convex combination
of pure quantum states $\rho _{j} (j=1,2,...,N)$, where $\rho _{j} ={\left|\phi_{j} \right\rangle}{\left\langle \phi _{j}  \right|} $, ${\left| \phi _{j}\right
\rangle} =\sum _{k=1}^{4} w_{jk} e^{i\theta _{jk}} {\left| b_{k}\right\rangle}$, $\theta
_{j1} =0$ (which is treated as a global phase) and $N$ is the number of pure quantum
states. Because $\bold{P}(\rho _{_{V} } )$ is at the boundary of the legal convex region, $\bold{P}(
\rho _{j} )$ should also at the boundary thereby with $w_{j3} =w_{j4} =0$ for any $j\in 1,2,...,N$. Then a straightforward computation leads to
\begin{equation} \label{ZEqnNum159336}
\rho _{j} ={\left| \phi _{j}  \right\rangle} {\left\langle \phi _{j}  \right|} =
\left(\begin{array}{cccc} {f_{j1} } & {0} & {0} & {f_{j2} -if_{j3} } \\ {0} & {0}
& {0} & {0} \\ {0} & {0} & {0} & {0} \\ {f_{j2} +if_{j3} } & {0} & {0} & {1-f_{j1}
} \end{array}\right),
\end{equation}
where
%\begin{equation} {ll} \label{19)}
\begin{subequations}
\begin{eqnarray}
&&f_{j1}={\left(w_{j1}^{2} +w_{j2}^{2} +2w_{j1} w_{j2} \cos (\theta
_{j2} )\right)/2} \\
&&{f_{j2} =\left(w_{j1}^{2} -w_{j2}^{2} \right)/2} \\
&&{f_{j3}=w_{j1} w_{j2} \sin (\theta _{j2} )}
\end{eqnarray}
\end{subequations}
%\end{equation}
Thus $\rho_{_{V}}$, as a convex combination of $\rho _{j} (j=1,2,...,N)$, can be
expressed as in above Eq. \eqref{ZEqnNum494785}.

Next, we prove that $f_{2} =f_{3} =0$ or $f_{2} =\sin (2\chi -2\psi
)=0$ is a necessary condition for the equation $\bold{P}_{V} (\rho )=(0,0,1)'$ to hold.
First, we prove $f_{2} =0$ if $\bold{P}(\rho _{_{V} } )=(0,0,1)'$. In fact, by Lemma \ref{lemma3},
we have $\bold{P}(\rho _{_{V} } )=(-2f_{2} ,-2f_{2} ,1)'$ thereby getting $f_{2} =0$ soon.
Supposing $\bold{P}(U)=(0,0,1)'$, by Lemma \ref{lemma2}, $\gamma _{1} $, $\gamma _{2} $ and $\varphi
_{0} $ of $U$ should be $(-1)^{n_{2} } \frac{\pi }{4} +k_{1} \pi $, 0 and 0 respectively,
where $n_{2} ,k_{1} \in \{ 0,1\} $. To get the necessary condition, we only need
to testify whether there is a unitary operator $V_{1} =V'$ with parameters being $\varphi
_{1} $, $\psi _{1} $ and $\chi _{1} $ as stated in Eq. \eqref{V} such that $\bold{P}_{V_{1}}(\rho_{_{V}})=\bold{P}_{V_{1}}(U)$. On the one hand,
by Lemma \ref{lemma3}, after calculation, we have
\begin{equation} \label{20)}
C_{33\_ V_{1} } (\rho )=\mathop{\cos }\nolimits^{2} (2\varphi _{1} )+2f_{3} \mathop{
\sin }\nolimits^{2} (2\varphi _{1} )\mathop{\sin }\nolimits^{2} (2\psi _{1} +2\chi
_{1} )
\end{equation}
And on the other hand, by Lemma \ref{lemma4}, we get
\begin{equation} \label{21)}
C_{33\_ V_{1} } (U)=\mathop{\cos }\nolimits^{2} (2\varphi _{1} )
\end{equation}
Thus by $C_{33\_ V_{1} } (\rho )=C_{33\_ V_{1} } (U)$, $2f_{3} \mathop{\sin }\nolimits^{2}
(2\varphi _{1} )\mathop{\sin }\nolimits^{2} (2\psi _{1} +2\chi _{1} )$ should be
0. Since $\mathop{\sin }\nolimits^{2} (2\varphi _{1} )\ne 0$($\bold{P}\ne (0,0,1)'$), we
get $\sin (2\psi _{1} +2\chi _{1} )=0$ or $f_{3} =0$. Then, by $V_{1} =V'$, $\varphi
_{1} =\varphi $, $\psi _{1} =-\psi $ and $\chi _{1} =\chi +\pi $. Thus, we finally
get the necessary condition is $f_{2} =f_{3} =0$ or $f_{2} =\sin (2\chi -2\psi )=0$.
\end{proof}

\section{\label{pt4}PROOF OF THEOREM 4}
\begin{proof}
Denote the new values of $\bold{P}$ for $\rho $ and $U$ are $\bold{P}_{-}
(\rho )$ and $\bold{P}_{-} (U)$. Obviously,
\begin{equation} \label{22)}
\bold{P}_{-} (\rho )=\bold{P}\left((V_{-} \otimes V_{{\rm +}} )'\rho \right)=\left(\begin{array}{c}
{0} \\ {0} \\ {-1} \end{array}\right).
\end{equation}
The sum of the three entries of $\bold{P}_{-} (\rho )$ is equal to -1, then $\bold{P}_{-}
(\rho )$ is in the plane $l(-1)$. For $\bold{P}_{-} (U)$, we first prove $\bold{P}_{-}(U)=\bold{P}(V_{{\rm +}}'UV_{-})$.
Supposing $\xi _{i} $ is one of the two spectral measures associated with an observable $\sigma
_{i} (i=1,2,3)$, we measure the qubit before and after unitary operation $U$. The
probability that the outcome of the measurement before unitary operation $U$ is $V_{-}{\left| \xi _{i}  \right\rangle} $  and the outcome of the measurement after unitary operation $U$ is $V_{+} {\left| \xi _{i}
\right\rangle} $ is
\begin{equation} \label{ZEqnNum539105}
{\left\langle \xi _{i}  \right|} V_{-} 'U'V_{+} {\left| \xi _{i}  \right\rangle}
{\left\langle \xi _{i}  \right|} V_{+} 'UV_{-} {\left| \xi _{i}  \right\rangle}
\end{equation}
where after the first measurement, the state of the qubit collapsed to $V_{-} {\left| \xi
_{i}\right\rangle} $. According to Eq. \eqref{ZEqnNum539105}, $V_{+}'UV_{-} $ can
be seen as a new $U$, then we get $\bold{P}_{-} (U)=\bold{P}(V_{+}'UV_{-} )$. By Lemma \ref{lemma2} and Lemma
\ref{lemma4}, a straightforward computation can soon gives
\begin{equation} \label{24)}
\bold{P}_{-} (U)=\left(\begin{array}{c} {0} \\ {0} \\ {-1} \end{array}\right).
\end{equation}
It is also in the plane $l(-1)$.
\end{proof}

\nocite{*}
\bibliography{mybib}

\begin{thebibliography}{24}%
\makeatletter
\providecommand \@ifxundefined [1]{%
 \@ifx{#1\undefined}
}%
\providecommand \@ifnum [1]{%
 \ifnum #1\expandafter \@firstoftwo
 \else \expandafter \@secondoftwo
 \fi
}%
\providecommand \@ifx [1]{%
 \ifx #1\expandafter \@firstoftwo
 \else \expandafter \@secondoftwo
 \fi
}%
\providecommand \natexlab [1]{#1}%
\providecommand \enquote  [1]{``#1''}%
\providecommand \bibnamefont  [1]{#1}%
\providecommand \bibfnamefont [1]{#1}%
\providecommand \citenamefont [1]{#1}%
\providecommand \href@noop [0]{\@secondoftwo}%
\providecommand \href [0]{\begingroup \@sanitize@url \@href}%
\providecommand \@href[1]{\@@startlink{#1}\@@href}%
\providecommand \@@href[1]{\endgroup#1\@@endlink}%
\providecommand \@sanitize@url [0]{\catcode `\\12\catcode `\$12\catcode
  `\&12\catcode `\#12\catcode `\^12\catcode `\_12\catcode `\%12\relax}%
\providecommand \@@startlink[1]{}%
\providecommand \@@endlink[0]{}%
\providecommand \url  [0]{\begingroup\@sanitize@url \@url }%
\providecommand \@url [1]{\endgroup\@href {#1}{\urlprefix }}%
\providecommand \urlprefix  [0]{URL }%
\providecommand \Eprint [0]{\href }%
\providecommand \doibase [0]{http://dx.doi.org/}%
\providecommand \selectlanguage [0]{\@gobble}%
\providecommand \bibinfo  [0]{\@secondoftwo}%
\providecommand \bibfield  [0]{\@secondoftwo}%
\providecommand \translation [1]{[#1]}%
\providecommand \BibitemOpen [0]{}%
\providecommand \bibitemStop [0]{}%
\providecommand \bibitemNoStop [0]{.\EOS\space}%
\providecommand \EOS [0]{\spacefactor3000\relax}%
\providecommand \BibitemShut  [1]{\csname bibitem#1\endcsname}%
\let\auto@bib@innerbib\@empty
%</preamble>
\bibitem [{\citenamefont {Renoirte}(1956)}]{renoirte1956hans}%
  \BibitemOpen
  \bibfield  {author} {\bibinfo {author} {\bibfnamefont {F.}~\bibnamefont
  {Renoirte}},\ }\href
  {https://www.persee.fr/doc/phlou_0035-3841_1956_num_54_43_4887_t1_0522_0000_2}
  {\bibfield  {journal} {\bibinfo  {journal} {Revue philosophique de Louvain}\
  }\textbf {\bibinfo {volume} {54}},\ \bibinfo {pages} {522} (\bibinfo {year}
  {1956})}\BibitemShut {NoStop}%
\bibitem [{\citenamefont {Pearl}(2009)}]{pearl2009causality}%
  \BibitemOpen
  \bibfield  {author} {\bibinfo {author} {\bibfnamefont {J.}~\bibnamefont
  {Pearl}},\ }\href
  {https://www.google.com/books?hl=zh-CN&lr=&id=f4nuexsNVZIC&oi=fnd&pg=PR1&dq=Causality&ots=y2EWTnvwkk&sig=riOLzAmSYa42c7afzuBF4RaxtqY}
  {\emph {\bibinfo {title} {Causality}}}\ (\bibinfo  {publisher} {Cambridge
  university press},\ \bibinfo {year} {2009})\BibitemShut {NoStop}%
\bibitem [{\citenamefont {Wood}\ and\ \citenamefont
  {Spekkens}(2015)}]{wood2015lesson}%
  \BibitemOpen
  \bibfield  {author} {\bibinfo {author} {\bibfnamefont {C.~J.}\ \bibnamefont
  {Wood}}\ and\ \bibinfo {author} {\bibfnamefont {R.~W.}\ \bibnamefont
  {Spekkens}},\ }\href
  {http://iopscience.iop.org/article/10.1088/1367-2630/17/3/033002/meta}
  {\bibfield  {journal} {\bibinfo  {journal} {New Journal of Physics}\ }\textbf
  {\bibinfo {volume} {17}},\ \bibinfo {pages} {033002} (\bibinfo {year}
  {2015})}\BibitemShut {NoStop}%
\bibitem [{\citenamefont {Evans}\ \emph {et~al.}(2012)\citenamefont {Evans},
  \citenamefont {Price},\ and\ \citenamefont {Wharton}}]{evans2012new}%
  \BibitemOpen
  \bibfield  {author} {\bibinfo {author} {\bibfnamefont {P.~W.}\ \bibnamefont
  {Evans}}, \bibinfo {author} {\bibfnamefont {H.}~\bibnamefont {Price}}, \ and\
  \bibinfo {author} {\bibfnamefont {K.~B.}\ \bibnamefont {Wharton}},\ }\href
  {https://academic.oup.com/bjps/article-abstract/64/2/297/1518728} {\bibfield
  {journal} {\bibinfo  {journal} {The British Journal for the Philosophy of
  Science}\ }\textbf {\bibinfo {volume} {64}},\ \bibinfo {pages} {297}
  (\bibinfo {year} {2012})}\BibitemShut {NoStop}%
\bibitem [{\citenamefont {Tucci}(1995)}]{tucci1995quantum}%
  \BibitemOpen
  \bibfield  {author} {\bibinfo {author} {\bibfnamefont {R.~R.}\ \bibnamefont
  {Tucci}},\ }\href@noop {} {\bibfield  {journal} {\bibinfo  {journal}
  {International Journal of Modern Physics B}\ }\textbf {\bibinfo {volume}
  {9}},\ \bibinfo {pages} {295} (\bibinfo {year} {1995})}\BibitemShut {NoStop}%
\bibitem [{\citenamefont {Laskey}(2007)}]{laskey2007quantum}%
  \BibitemOpen
  \bibfield  {author} {\bibinfo {author} {\bibfnamefont {K.~B.}\ \bibnamefont
  {Laskey}},\ }in\ \href@noop {} {\emph {\bibinfo {booktitle} {AAAI Spring
  Symposium: Quantum Interaction}}}\ (\bibinfo {year} {2007})\ pp.\ \bibinfo
  {pages} {142--149}\BibitemShut {NoStop}%
\bibitem [{\citenamefont {Pienaar}\ and\ \citenamefont
  {Brukner}(2015)}]{pienaar2015graph}%
  \BibitemOpen
  \bibfield  {author} {\bibinfo {author} {\bibfnamefont {J.}~\bibnamefont
  {Pienaar}}\ and\ \bibinfo {author} {\bibfnamefont {{\v{C}}.}~\bibnamefont
  {Brukner}},\ }\href@noop {} {\bibfield  {journal} {\bibinfo  {journal} {New
  Journal of Physics}\ }\textbf {\bibinfo {volume} {17}},\ \bibinfo {pages}
  {073020} (\bibinfo {year} {2015})}\BibitemShut {NoStop}%
\bibitem [{\citenamefont {Fritz}(2016)}]{fritz2016beyond}%
  \BibitemOpen
  \bibfield  {author} {\bibinfo {author} {\bibfnamefont {T.}~\bibnamefont
  {Fritz}},\ }\href@noop {} {\bibfield  {journal} {\bibinfo  {journal}
  {Communications in Mathematical Physics}\ }\textbf {\bibinfo {volume}
  {341}},\ \bibinfo {pages} {391} (\bibinfo {year} {2016})}\BibitemShut
  {NoStop}%
\bibitem [{\citenamefont {Leifer}(2006)}]{leifer2006quantum}%
  \BibitemOpen
  \bibfield  {author} {\bibinfo {author} {\bibfnamefont {M.~S.}\ \bibnamefont
  {Leifer}},\ }\href
  {https://journals.aps.org/pra/abstract/10.1103/PhysRevA.74.042310} {\bibfield
   {journal} {\bibinfo  {journal} {Physical Review A}\ }\textbf {\bibinfo
  {volume} {74}},\ \bibinfo {pages} {042310} (\bibinfo {year}
  {2006})}\BibitemShut {NoStop}%
\bibitem [{\citenamefont {Henson}\ \emph {et~al.}(2014)\citenamefont {Henson},
  \citenamefont {Lal},\ and\ \citenamefont {Pusey}}]{henson2014theory}%
  \BibitemOpen
  \bibfield  {author} {\bibinfo {author} {\bibfnamefont {J.}~\bibnamefont
  {Henson}}, \bibinfo {author} {\bibfnamefont {R.}~\bibnamefont {Lal}}, \ and\
  \bibinfo {author} {\bibfnamefont {M.~F.}\ \bibnamefont {Pusey}},\ }\href@noop
  {} {\bibfield  {journal} {\bibinfo  {journal} {New Journal of Physics}\
  }\textbf {\bibinfo {volume} {16}},\ \bibinfo {pages} {113043} (\bibinfo
  {year} {2014})}\BibitemShut {NoStop}%
\bibitem [{\citenamefont {Allen}\ \emph {et~al.}(2017)\citenamefont {Allen},
  \citenamefont {Barrett}, \citenamefont {Horsman}, \citenamefont {Lee},\ and\
  \citenamefont {Spekkens}}]{allen2017quantum}%
  \BibitemOpen
  \bibfield  {author} {\bibinfo {author} {\bibfnamefont {J.-M.~A.}\
  \bibnamefont {Allen}}, \bibinfo {author} {\bibfnamefont {J.}~\bibnamefont
  {Barrett}}, \bibinfo {author} {\bibfnamefont {D.~C.}\ \bibnamefont
  {Horsman}}, \bibinfo {author} {\bibfnamefont {C.~M.}\ \bibnamefont {Lee}}, \
  and\ \bibinfo {author} {\bibfnamefont {R.~W.}\ \bibnamefont {Spekkens}},\
  }\href {https://journals.aps.org/prx/abstract/10.1103/PhysRevX.7.031021}
  {\bibfield  {journal} {\bibinfo  {journal} {Physical Review X}\ }\textbf
  {\bibinfo {volume} {7}},\ \bibinfo {pages} {031021} (\bibinfo {year}
  {2017})}\BibitemShut {NoStop}%
\bibitem [{\citenamefont {Giarmatzi}\ and\ \citenamefont
  {Costa}(2018)}]{giarmatzi2018quantum}%
  \BibitemOpen
  \bibfield  {author} {\bibinfo {author} {\bibfnamefont {C.}~\bibnamefont
  {Giarmatzi}}\ and\ \bibinfo {author} {\bibfnamefont {F.}~\bibnamefont
  {Costa}},\ }\href {https://www.nature.com/articles/s41534-018-0062-6}
  {\bibfield  {journal} {\bibinfo  {journal} {npj Quantum Information}\
  }\textbf {\bibinfo {volume} {4}},\ \bibinfo {pages} {17} (\bibinfo {year}
  {2018})}\BibitemShut {NoStop}%
\bibitem [{\citenamefont {Costa}\ and\ \citenamefont
  {Shrapnel}(2016)}]{costa2016quantum}%
  \BibitemOpen
  \bibfield  {author} {\bibinfo {author} {\bibfnamefont {F.}~\bibnamefont
  {Costa}}\ and\ \bibinfo {author} {\bibfnamefont {S.}~\bibnamefont
  {Shrapnel}},\ }\href
  {http://iopscience.iop.org/article/10.1088/1367-2630/18/6/063032/meta}
  {\bibfield  {journal} {\bibinfo  {journal} {New Journal of Physics}\ }\textbf
  {\bibinfo {volume} {18}},\ \bibinfo {pages} {063032} (\bibinfo {year}
  {2016})}\BibitemShut {NoStop}%
\bibitem [{\citenamefont {Fitzsimons}\ \emph {et~al.}(2015)\citenamefont
  {Fitzsimons}, \citenamefont {Jones},\ and\ \citenamefont
  {Vedral}}]{fitzsimons2013quantum}%
  \BibitemOpen
  \bibfield  {author} {\bibinfo {author} {\bibfnamefont {J.~F.}\ \bibnamefont
  {Fitzsimons}}, \bibinfo {author} {\bibfnamefont {J.~A.}\ \bibnamefont
  {Jones}}, \ and\ \bibinfo {author} {\bibfnamefont {V.}~\bibnamefont
  {Vedral}},\ }\href@noop {} {\bibfield  {journal} {\bibinfo  {journal}
  {Scientific reports}\ }\textbf {\bibinfo {volume} {5}},\ \bibinfo {pages}
  {18281} (\bibinfo {year} {2015})}\BibitemShut {NoStop}%
\bibitem [{\citenamefont {Ried}\ \emph {et~al.}(2015)\citenamefont {Ried},
  \citenamefont {Agnew}, \citenamefont {Vermeyden}, \citenamefont {Janzing},
  \citenamefont {Spekkens},\ and\ \citenamefont {Resch}}]{ried2015quantum}%
  \BibitemOpen
  \bibfield  {author} {\bibinfo {author} {\bibfnamefont {K.}~\bibnamefont
  {Ried}}, \bibinfo {author} {\bibfnamefont {M.}~\bibnamefont {Agnew}},
  \bibinfo {author} {\bibfnamefont {L.}~\bibnamefont {Vermeyden}}, \bibinfo
  {author} {\bibfnamefont {D.}~\bibnamefont {Janzing}}, \bibinfo {author}
  {\bibfnamefont {R.~W.}\ \bibnamefont {Spekkens}}, \ and\ \bibinfo {author}
  {\bibfnamefont {K.~J.}\ \bibnamefont {Resch}},\ }\href
  {https://www.nature.com/articles/nphys3266} {\bibfield  {journal} {\bibinfo
  {journal} {Nature Physics}\ }\textbf {\bibinfo {volume} {11}},\ \bibinfo
  {pages} {414} (\bibinfo {year} {2015})}\BibitemShut {NoStop}%
\bibitem [{\citenamefont {{K{\"u}bler}}\ and\ \citenamefont
  {{Braun}}(2018)}]{kubler2018two}%
  \BibitemOpen
  \bibfield  {author} {\bibinfo {author} {\bibfnamefont {J.~M.}\ \bibnamefont
  {{K{\"u}bler}}}\ and\ \bibinfo {author} {\bibfnamefont {D.}~\bibnamefont
  {{Braun}}},\ }\href {\doibase 10.1088/1367-2630/aad612} {\bibfield  {journal}
  {\bibinfo  {journal} {New Journal of Physics}\ }\textbf {\bibinfo {volume}
  {20}},\ \bibinfo {pages} {083015} (\bibinfo {year} {2018})}\BibitemShut
  {NoStop}%
\bibitem [{\citenamefont {Chiribella}\ \emph {et~al.}(2010)\citenamefont
  {Chiribella}, \citenamefont {D'Ariano},\ and\ \citenamefont
  {Perinotti}}]{chiribella2010probabilistic}%
  \BibitemOpen
  \bibfield  {author} {\bibinfo {author} {\bibfnamefont {G.}~\bibnamefont
  {Chiribella}}, \bibinfo {author} {\bibfnamefont {G.~M.}\ \bibnamefont
  {D'Ariano}}, \ and\ \bibinfo {author} {\bibfnamefont {P.}~\bibnamefont
  {Perinotti}},\ }\href {\doibase 10.1103/PhysRevA.81.062348} {\bibfield
  {journal} {\bibinfo  {journal} {Phys. Rev. A}\ }\textbf {\bibinfo {volume}
  {81}},\ \bibinfo {pages} {062348} (\bibinfo {year} {2010})}\BibitemShut
  {NoStop}%
\bibitem [{\citenamefont {Hu}\ and\ \citenamefont
  {Hou}(2018)}]{hu2018discrimination}%
  \BibitemOpen
  \bibfield  {author} {\bibinfo {author} {\bibfnamefont {M.}~\bibnamefont
  {Hu}}\ and\ \bibinfo {author} {\bibfnamefont {Y.}~\bibnamefont {Hou}},\
  }\href {https://journals.aps.org/pra/abstract/10.1103/PhysRevA.97.062125}
  {\bibfield  {journal} {\bibinfo  {journal} {Physical Review A}\ }\textbf
  {\bibinfo {volume} {97}},\ \bibinfo {pages} {062125} (\bibinfo {year}
  {2018})}\BibitemShut {NoStop}%
\bibitem [{\citenamefont {Rivas}\ \emph {et~al.}(2014)\citenamefont {Rivas},
  \citenamefont {Huelga},\ and\ \citenamefont {Plenio}}]{rivas2014quantum}%
  \BibitemOpen
  \bibfield  {author} {\bibinfo {author} {\bibfnamefont {{\'{A}}.}~\bibnamefont
  {Rivas}}, \bibinfo {author} {\bibfnamefont {S.~F.}\ \bibnamefont {Huelga}}, \
  and\ \bibinfo {author} {\bibfnamefont {M.~B.}\ \bibnamefont {Plenio}},\
  }\href {\doibase 10.1088/0034-4885/77/9/094001} {\bibfield  {journal}
  {\bibinfo  {journal} {Reports on Progress in Physics}\ }\textbf {\bibinfo
  {volume} {77}},\ \bibinfo {pages} {094001} (\bibinfo {year}
  {2014})}\BibitemShut {NoStop}%
\bibitem [{\citenamefont {Laine}\ \emph {et~al.}(2010)\citenamefont {Laine},
  \citenamefont {Piilo},\ and\ \citenamefont {Breuer}}]{laine2010measure}%
  \BibitemOpen
  \bibfield  {author} {\bibinfo {author} {\bibfnamefont {E.-M.}\ \bibnamefont
  {Laine}}, \bibinfo {author} {\bibfnamefont {J.}~\bibnamefont {Piilo}}, \ and\
  \bibinfo {author} {\bibfnamefont {H.-P.}\ \bibnamefont {Breuer}},\ }\href
  {\doibase 10.1103/PhysRevA.81.062115} {\bibfield  {journal} {\bibinfo
  {journal} {Phys. Rev. A}\ }\textbf {\bibinfo {volume} {81}},\ \bibinfo
  {pages} {062115} (\bibinfo {year} {2010})}\BibitemShut {NoStop}%
\bibitem [{\citenamefont {Chiribella}\ \emph {et~al.}(2013)\citenamefont
  {Chiribella}, \citenamefont {D{\'A}riano},\ and\ \citenamefont
  {Roetteler}}]{chiribella2013identification}%
  \BibitemOpen
  \bibfield  {author} {\bibinfo {author} {\bibfnamefont {G.}~\bibnamefont
  {Chiribella}}, \bibinfo {author} {\bibfnamefont {G.~M.}\ \bibnamefont
  {D{\'A}riano}}, \ and\ \bibinfo {author} {\bibfnamefont {M.}~\bibnamefont
  {Roetteler}},\ }\href {\doibase 10.1088/1367-2630/15/10/103019} {\bibfield
  {journal} {\bibinfo  {journal} {New Journal of Physics}\ }\textbf {\bibinfo
  {volume} {15}},\ \bibinfo {pages} {103019} (\bibinfo {year}
  {2013})}\BibitemShut {NoStop}%
\bibitem [{\citenamefont {Chiribella}(2012)}]{chiribella2012perfect}%
  \BibitemOpen
  \bibfield  {author} {\bibinfo {author} {\bibfnamefont {G.}~\bibnamefont
  {Chiribella}},\ }\href {\doibase 10.1103/PhysRevA.86.040301} {\bibfield
  {journal} {\bibinfo  {journal} {Phys. Rev. A}\ }\textbf {\bibinfo {volume}
  {86}},\ \bibinfo {pages} {040301} (\bibinfo {year} {2012})}\BibitemShut
  {NoStop}%
\bibitem [{\citenamefont {Shrapnel}(2015)}]{shrapnel2015discovering}%
  \BibitemOpen
  \bibfield  {author} {\bibinfo {author} {\bibfnamefont {S.}~\bibnamefont
  {Shrapnel}},\ }\href@noop {} {\bibfield  {journal} {\bibinfo  {journal} {The
  British Journal for the Philosophy of Science}\ } (\bibinfo {year}
  {2015})}\BibitemShut {NoStop}%
\bibitem [{\citenamefont {Pienaar}(2017)}]{pienaar2017causality}%
  \BibitemOpen
  \bibfield  {author} {\bibinfo {author} {\bibfnamefont {J.}~\bibnamefont
  {Pienaar}},\ }\href@noop {} {\bibfield  {journal} {\bibinfo  {journal}
  {Physics}\ }\textbf {\bibinfo {volume} {10}},\ \bibinfo {pages} {86}
  (\bibinfo {year} {2017})}\BibitemShut {NoStop}%
\end{thebibliography}%

\end{document}